\documentclass[fleqn, usenatbib]{mnras}

\usepackage{newtxtext, newtxmath}

\usepackage[T1]{fontenc}

\DeclareRobustCommand{\VAN}[3]{#2}
\let\VANthebibliography\thebibliography
\def\thebibliography{\DeclareRobustCommand{\VAN}[3]{##3}\VANthebibliography}

\usepackage{graphicx} 
\usepackage{amsmath} 

\title[PBI-enhanced outflow-based corona]{Unifying the X-ray coronae and ultra-fast
outflows: a PBI-enhanced outflow-based corona model for the inner accretion disc}

\author[H. Xu]{ Haichao Xu
\thanks{E-mail: haichao\_xu@zju.edu.cn (HX)}
\\
Institute for Astronomy, School of Physics, Zhejiang University, 866 Yuhangtang Road,
Hangzhou 310058, China }

\date{Accepted 2026 March 24. Received 2026 March 11; in original form 2026 January 23}

\pubyear{\the\year{}}

\begin{document}
	\label{firstpage} \pagerange{\pageref{firstpage}--\pageref{lastpage}}
	\maketitle

	\begin{abstract}
		{The fact that luminous X-ray coronae and Ultra-Fast Outflows (UFOs) are both inferred to originate from the innermost regions of active galactic nuclei (AGNs) suggests a deep physical connection between them.}
		However, standard magnetic buoyancy models struggle to transport
		sufficient energy through the radiation-pressure-dominated inner disc to
		sustain both the phenomena, creating a theoretical energy deficit. In
		this work, we propose an outflow-based model with energy transport enhanced
		by the Photon Bubble Instability (PBI) in the inner region. By coupling
		this enhanced energy supply with the
		{MHD turbulence-driven mass-loading mechanism} appropriate for weakly
		magnetized standard discs, we solve the dynamical and thermodynamic structure
		of the corona. We find that the model can successfully launch high speed
		winds matching observed UFO kinematics provided the mechanical
		acceleration efficiency is high ($f_{\rm acc}\gtrsim 0.5$). Furthermore,
		the model naturally reproduces the observed spectral evolution found in
		AGN coronae: as the accretion rate increases, the corona becomes
		optically thicker and cooler and produces a softer spectrum. Our results
		support an extended slab-like coronal geometry and suggest that UFOs and
		X-ray coronae in the inner discs are manifestations of the same magnetic
		activity.
	\end{abstract}

	\begin{keywords}
		accretion, accretion discs – galaxies: active – X-rays: galaxies.
	\end{keywords}



	\section{Introduction}

	Accretion onto supermassive black hole (BH) in active galactic nuclei (AGNs)
	is widely regarded as the primary power source for these luminous objects
	\citep[e.g., ][hereafter SS73]{lynden-bellGalacticNucleiCollapsed1969a, ss73}.
	In the standard thin disc picture, gravitational energy is dissipated and
	radiated efficiently within an optically thick and geometrically thin disc, whose
	thermal emission peaks in the optical/ultraviolet (UV) band (e.g., SS73).
	However, the observed continua commonly exhibit an additional luminous X-ray
	component, which is often attributed to inverse Compton scattering (IC) of
	disc photons by a hot, tenuous electron plasma \citep[e.g.,][]{1976ApJ...204..187S, 1980A&A....86..121S}.
	This plasma is conventionally referred to as the corona and forms the
	cornerstone of two-phase disc-corona scenarios for producing the hard X-ray
	power-law emission \citep[e.g.,][]{1977ApJ...218..247L, GS79, HM91, HM93, 1994ApJ...436..599S}.
	Despite extensive progress on accretion disc theory and phenomenology, the
	physical nature of the corona still remains unsettled.

	Although the existence of corona in AGNs is well established, their heating mechanism
	and precise geometry remain matters of debate. A variety of corona models
	based on different geometric and physical processes have been proposed, including
	extended slab (“sandwich”) corona overlying a thin disc\citep[e.g.,][]{HM91, HM93},
	truncated-disc/hot-inner-flow scenarios \citep[e.g.,][]{1997ApJ...489..865E, 2007A&ARv..15....1D}
	and compact axial sources idealized as lampposts or jet-based structures \citep[e.g.,][]{1991A&A...247...25M, 2013ApJ...768..146G, 2005ApJ...635.1203M}.

	Recently, accumulating observational evidence favours the extended geometry.
	Analyses of the compactness-temperature relationship in large AGN samples suggest
	that the X-ray emitting region is likely geometrically extended rather than compact
	point-like \citep{2015MNRAS.451.4375F, 2018MNRAS.480.1819R}. Crucially, the
	advent of X-ray polarimetry with Imaging X-ray Polarimetry Explorer (IXPE)
	has provided a decisive diagnostic \citep[e.g.,][]{2022JATIS...8b6002W}. Polarimetric
	observations of both X-ray binaries (XRBs) and AGNs consistently reveal polarization
	vectors aligned with the jet axis (or perpendicular to the disc plane), a
	signature characteristic of an extended scattering geometry \citep[e.g.][]{2022Sci...378..650K, 2023MNRAS.523.4468G, 2024ApJ...974..101S}.
	Furthermore, recent modeling of Cyg X-1 indicates that the sandwich corona,
	rather than the truncated disc model, offers a superior explanation for the observed
	polarization angle and degree \citep[e.g.,][]{2025PASJ..tmp..125Z}.
	Motivated by these compelling observational developments, we adopt the slab coronal
	geometry as the baseline framework for our investigation.

	Multiple observational diagnostics indicate that the X-ray emission in XRBs and
	AGNs is produced in a compact region, with characteristic sizes often
	constrained within a few tens of gravitational radii
	$R_{\rm g}= G M_{\rm BH}/c^{2}$ (e.g., via gravitational microlensing \citep[e.g.][]{2009ApJ...693..174C, 2013ApJ...769...53M, 2016AN....337..356C},
	X-ray reverberation \citep[e.g.][]{2013MNRAS.431.2441D, 2014A&ARv..22...72U},
	and radio spectral analysis\citep[e.g.,][]{2024A&A...690A.232S, 2025MNRAS.539..808M}).
	Therefore, if the cold thin disc extends down to the innermost stable circular
	orbit (ISCO), the dominant coronal emission is expected to be coupled
	primarily to the inner disc, which corresponds to the radiation-pressure-dominated
	zone in the standard accretion disc framework (e.g., SS73). This motivates a
	careful treatment of the disc-corona energy transport in the inner disc
	region.

	In many slab corona scenarios, the coronal power is assumed to originate from
	magnetorotational instability (MRI) in weakly magnetized accretion discs \citep[e.g.,][]{BH91, MS00}.
	The magnetic flux can rise buoyantly into the corona and dissipate through
	reconnection or wave dissipation to heat the tenuous plasma \citep[e.g.,][hereafter
	MF02 and C09]{GS79, 1984ApJ...277..312S, 1998MNRAS.299L..15D, MF02, C09}.
	However, semi-analytic buoyancy-powered disc-corona models often struggle to
	yield a larger coronal power fraction $f$ in the radiation-pressure-dominated
	region; for instance, MF02 argued that the magnetic energy density is tied to
	gas pressure, and $f$ is suppressed where radiation pressure dominates.

	On the other hand, in the hard state of XRBs the Comptonized X-ray component
	can carry a substantial fraction of the radiative output \citep[e.g.,][]{remillardXRayPropertiesBlackHole2006},
	implying an efficient disc-corona coupling in at least some conditions; yet in
	frameworks such as MF02 and C09 the inferred $f$ depends sensitively on the viscosity
	parameter $\alpha$, so that producing a bright corona may require unusually
	large $\alpha$ \citep[e.g.,][also see MF02 and C09]{2025MNRAS.544.1748X},
	seemingly in tension with typical MRI-turbulence levels found in simulations
	and observations \citep[e.g.,][]{2007MNRAS.376.1740K}. This suggests that an
	additional mechanism may be missing in the radiation-pressure-dominated region,
	capable of enhancing vertical magnetic energy transport in their theoretical
	models.

	Fast outflows/winds from the inner disc region have been detected in many
	AGNs and XRBs, including ultra-fast outflows (UFOs) in AGNs \citep[e.g.,][]{2010A&A...521A..57T, 2012MNRAS.422L...1T, 2013MNRAS.430...60G}
	and highly ionized disc winds in XRBs \citep[e.g.,][]{2006Natur.441..953M, 2009Natur.458..481N},
	with launching region in some sources also inferred to be relatively compact
	\citep[e.g.,][]{2012MNRAS.422L...1T, 2013MNRAS.430.1102T, 2015MNRAS.451.4169G, 2025arXiv251206077L}.
	If the X-ray emission is produced predominantly by a compact inner corona, a
	strong spatial coincidence between the corona and these outflows is
	indicated. The corona may naturally act as an energetic and dynamical base for
	these outflows. Although there are other mechanisms that can also generate
	similar fast outflows, for example, magnetocentrifugal acceleration
	mechanism \citep[e.g.,][]{1982MNRAS.199..883B}. However, it requires a large-scale
	ordered strong poloidal magnetic field whose ubiquity remains uncertain
	\citep[e.g.,][]{1994MNRAS.267..235L}. In this context, treating the corona
	as part of the outflow base implies an additional energetic requirement: powering
	the intense X-ray emission while simultaneously supplying the kinetic power to
	launch and accelerate the wind. Notably, MF02 already pointed out that energetically
	dominant coronae are natural sites for launching powerful jets/outflows,
	linking a “bright corona” to the outflow energetics.

	In the radiation-pressure-dominated region, where magnetic pressure is comparable
	to or exceeds gas pressure, the coupling of radiative diffusion to
	magnetized plasma can trigger the photon-bubble instability (PBI) \citep[e.g.,][]{1992ApJ...388..561A, 1998MNRAS.297..929G, 2001ApJ...551..897B, 2001ApJ...553..987B, 2003ApJ...596..509B}.
	{Although global radiation MHD simulations have historically failed to capture PBI \citep[e.g.,][]{2007ApJ...664.1057B, 2014ApJ...796..106J}, more recent studies have confirmed the existence of the PBI in laboratory cold atom clouds \citep[e.g.,][]{2021NatCo..12.3240G, 2022Atoms..10...45R} and neutron star environments \citep[e.g.,][]{2021MNRAS.508..617Z, 2022MNRAS.515.4371Z}. If the PBI also exists in BH accretion disc environments, its nonlinear development produces strong density inhomogeneities and may enhance vertical energy transport. This mechanism offers a natural physical pathway to sustain a large coronal fraction $f$ in the radiation-dominated inner disc.}
	We assess whether this enhanced corona can energetically support the observed
	outflows by adopting a parametric wind model. We treat the wind primarily as
	a site for mass and energy loss to test the robustness of the coronal power
	budget; the detailed magnetohydrodynamic (MHD) launching structure and the full
	vertical closure will be addressed in a subsequent study (Paper II).

	The paper is organized as follows. In Sec. \ref{sec:buoyancy_pbi}, we
	calculate the typical rising velocity of the magnetic flux tube and describe
	the physical mechanism of magnetic buoyancy enhanced by PBI. Sec.
	\ref{sec:global_disc} presents the global accretion disc solutions
	incorporating mass loss from the outflow. In Sec. \ref{sec:model}, we construct
	a simplified outflow-based coronal model to explore the dynamical and thermodynamic
	properties of the corona. In Sec. \ref{sec:varyingmdot}, we investigated the
	changes of the outflow and the corona when the accretion rate varies. Finally,
	we discuss the implications and summarize our findings in Sec. \ref{sec:disscussion}.

	\section{Vertical energy transport based on the magnetic buoyancy}
	\label{sec:buoyancy_pbi}

	\subsection{Buoyant Rise of Magnetic Flux Tubes}

	{In this subsection, we briefly review the standard magnetic buoyancy framework established in the literature (e.g., MF02; C09) and provide a simple derivation based on buoyancy-drag balance to set the physical stage for our model.}

	In the classical accretion disc picture, the angular momentum transport is governed
	by the Maxwell stress associated with MHD turbulence driven by the MRI \citep[e.g.][]{BH91, 1998RvMP...70....1B}.
	Among the stress components, the $r\phi$ component is the principal
	contributor to outward angular-momentum flux, which can be written as
	\begin{equation}
		\tau_{r\phi}= -\left\langle\frac{B_{r}B_{\phi}}{4\pi}\right\rangle,
	\end{equation}
	where $B_{r}$ and $B_{\phi}$ denote the radial and azimuthal magnetic-field components,
	and angle brackets represent appropriate space-time averages.

	The stress is typically parameterized using the $\alpha$ prescription. In the
	standard model of SS73, the stress is relative to the local total pressure, i.e.
	\begin{equation}
		\tau_{r\phi}= \alpha P_{\rm tot}= \alpha (P_{\rm gas}+ P_{\rm rad}),
	\end{equation}
	where $P_{\rm gas}$ and $P_{\rm rad}$ denote the gas pressure and radiation
	pressure, respectively. Although this assumption has been quite successful
	in many cases, the region dominated by radiation pressure is proved to be viscously
	and thermally unstable \citep[e.g.,][]{1974ApJ...187L...1L, 1976MNRAS.175..613S}.
	Alternatively, physically motivated arguments suggest that the magnetic field
	saturation may be limited by the gas pressure in the radiation-dominated region,
	leading to modified prescriptions
	$\tau_{r\phi}= \alpha \sqrt{P_{\rm gas}P_{\rm tot}}$ \citep[e.g.,][]{1984ApJ...287..761T}
	or $\tau_{r\phi}=\alpha P_{\rm gas}$ \citep[e.g.,][]{1984ApJ...277..312S}. As
	pointed by C09, different forms of viscous stress have a very significant
	impact on the properties of the corona region.

	MRI turbulence typically produces a predominantly azimuthal field in the
	saturated state, motivating the approximation \citep[e.g.,][]{MS00, 2014ApJ...796..106J}
	\begin{equation}
		P_{\rm mag}\approx\left\langle\frac{B_{\phi}^{2}}{8\pi}\right\rangle.
	\end{equation}
	Moreover, local MRI simulations indicate a comparatively robust scaling
	between viscosity stress and magnetic pressure \citep[e.g.,][]{MS00, 2008NewA...13..244B, 2014ApJ...796..106J},
	often quantified by
	\begin{equation}
		C_{\rm mag}=\frac{P_{\rm mag}}{\tau_{r\phi}},
	\end{equation}
	where $C_{\rm mag}$ is typically between 2 and 5, with details depending on
	numerical configuration. In our calculation, we take $C_{\rm mag}=3$. In disc-corona
	models following MF02/C09, it is common to adopt an effective identification
	$\tau_{r\phi}\approx P_{\rm mag}$. This approximation should be interpreted as
	an effective parametrization, since uncertainties typically enter
	multiplicatively and can be absorbed into an efficiency factor without
	altering the main, testable trends.

	In MF02/C09 and subsequent reconnection-heated slab corona frameworks, the
	buoyant rise speed of magnetic flux is often assumed to scale with the local
	Alfvén speed,
	\begin{equation}
		v_{\rm rise}= bv_{\rm A},
	\end{equation}
	where $b$ is dimensionless coefficient. However, such a proportionality prescription
	provides limited physical guidance on the magnitude and parameter dependence
	of $b$ \citep[e.g.,][]{2012ApJ...761..109Y}. In fact, we can derive the characteristic
	rise speed from a buoyancy-drag balance \citep[e.g.,][]{1975ApJ...198..205P, 1979ApJ...230..914P, 1984ApJ...277..312S},
	enabling us to rewrite $b$ in terms of more transparent geometric and
	thermodynamic quantities and setting the stage for incorporating PBI-induced
	modifications.

	Magnetic fields can be efficiently amplified in accretion discs via MRI, with
	a maximum linear growth rate of about $\Gamma_{\rm MRI}\sim 0.75\Omega_{\rm
	K}$ \citep[e.g.,][]{BH91}. We therefore adopt the characteristic MRI
	wavelength $\lambda_{\rm MRI}$ as an estimate of the coherence scale of
	buoyant magnetic structures, and approximate an emerging magnetic flux tube as
	a cylindrical flux tube with radius
	\begin{equation}
		a\approx\frac{\lambda_{\rm MRI}}{2}.
	\end{equation}

	Once a large-scale magnetic flux tube emerges, it must satisfy a pressure balance
	between the interior and exterior of the tube:
	\begin{equation}
		P_{\rm mag}+ P_{\rm tot, in}= P_{\rm tot}, \label{eq:pressure_balance}
	\end{equation}
	{where $P_{\rm tot}$ is the local total pressure of the disc and $P_{\rm tot, in}$ is the total pressure inside the tube.}
	Since $P_{\rm tot, in}<P_{\rm tot}$, the material density inside the tube is
	lower than that of the surrounding disc. As a result, Parker instability is excited,
	and the density difference causes the tube to experience an upward buoyant
	force \citep[e.g.,][]{1955ApJ...121..491P, 1966ApJ...145..811P}. The thermal
	pressure deficit across the tube is $\Delta P = P_{\rm mag}$.

	The plasma $\beta$ parameter is defined as
	\begin{equation}
		\beta = \frac{P_{\rm tot}}{P_{\rm mag}}= \frac{P_{\rm gas}+ P_{\rm rad}}{P_{\rm
		mag}}.
	\end{equation}
	In the accretion disc, $\beta \gg 1$ always holds.

	{The buoyant force per unit tube length for a given density difference on the magnetic tube is \begin{equation}F_{\rm buoy}= \pi a^{2}\Delta \rho g_{\rm d}, \label{eq:buoy_original}\end{equation} where $g_{\rm d}$ is the vertical acceleration due to gravity in the accretion disc. Therefore, to determine the buoyant force, we must evaluate the density deficit across the flux tube.}

	{In the gas-pressure-dominated region, previous studies have demonstrated that heat exchange is extremely efficient, allowing the flux tube to maintain thermal equilibrium with its surroundings \citep[e.g.,][]{1984ApJ...277..312S}. Therefore, it is typical to approximate $P_{\rm tot}\approx P_{\rm gas}$ in (\ref{eq:pressure_balance}) and adopt an isothermal assumption \citep[e.g.,][]{1975ApJ...198..205P}. The density deficit can then be calculated from the pressure difference: \begin{equation}\Delta \rho = \frac{P_{\rm mag}}{P_{\rm tot}}\rho_{\rm d}= \frac{v_{\rm A}^{2}}{2 c_{\rm s}^{2}}\rho_{\rm d},\end{equation} where $v_{\rm A}= \sqrt{2P_{\rm mag}/\rho_{\rm d}}$ is the Alfv\'en speed and $c_{\rm s}= \sqrt{P_{\rm tot}/ \rho_{\rm d}}$ is the isothermal sound speed.}

	{However, in the radiation-pressure-dominated region, applying this strict isothermal assumption leads to a physical contradiction. Global radiation-MHD simulations demonstrate that the magnetic pressure can readily reach levels comparable to or even exceeding the local gas pressure, i.e., $P_{\rm mag}\gtrsim P_{\rm gas}$ \citep[e.g.,][]{2014ApJ...796..106J, 2019ApJ...880...67J, 2016MNRAS.459.4397S}, which is consistent with the expectations of the standard accretion disc model. Consequently, the pressure deficit cannot be balanced solely by a reduction in gas pressure under the isothermal assumption. The internal temperature must drop relative to the external environment to allow the radiation pressure to compensate for the strong magnetic pressure.}\footnote{We
	note that \citet{1984ApJ...277..312S} adopted a modified viscosity prescription,
	$\tau_{r\phi}= \alpha P_{\rm gas}$, which inherently restricts the magnetic pressure
	so that it cannot exceed the gas pressure ($P_{\rm mag}\lesssim P_{\rm gas}$).
	Under this specific restriction, their strict isothermal assumption can still
	yield a mathematically viable rising speed. However, because modern global simulations
	demonstrate that $P_{\rm mag}$ can significantly exceed $P_{\rm gas}$ in the
	radiation-dominated inner disc, this strict isothermal derivation cannot be
	universally generalized to such strong-field environments.}

	{To model this non-isothermal state without solving the full radiative transfer equations during the dynamical rise, we introduce an effective polytropic equation of state to close the equations: \begin{equation}P_{\rm tot}= K \rho^{\gamma},\end{equation} where $\gamma$ is the polytropic index. For a radiation-dominated gas where radiation is partially trapped during the ascent, we adopt $\gamma = 4/3$ as a reasonable baseline approximation \citep[e.g.,][]{1990sse..book.....K, 2008bhad.book.....K}. Under this relation, the density deficit is correlated with the pressure deficit as: \begin{equation}\Delta \rho \approx \frac{\rm d \rho}{{\rm d}P_{\rm tot}}\Delta P = \frac{v_{\rm A}^{2}}{2\gamma c_{\rm s}^{2}}\rho_{\rm d}. \label{eq:deltaRho_final}\end{equation} We emphasize that this simple polytropic description is just an idealization. However, the exact value of $\gamma$ only alters the rising speed by a factor of order unity. More importantly, as we will demonstrate in Sec. \ref{sec:pbi}, once the Photon Bubble Instability (PBI) develops, the flux tube rapidly evacuates, rendering the specific choice of the initial polytropic index practically irrelevant to the terminal velocity.}

	In (\ref{eq:buoy_original}), the buoyant force on the magnetic tube is
	greater at higher altitudes. Therefore, for magnetic tubes that rise to a
	sufficient height in the accretion disc, we can estimate the buoyant force on
	the magnetic tube with (\ref{eq:deltaRho_final}):
	\begin{equation}
		F_{\rm buoy}\approx \pi a^{2}\left(\rho_{\rm d}\frac{v_{\rm A}^{2}}{2\gamma
		c_{\rm s}^{2}}\right )\Omega_{\rm K}^{2}H_{\rm d}, \label{eq:buoy}
	\end{equation}
	where $H_{\rm d}= c_{\rm s}/ \Omega_{\rm K}$ is the scale height of the
	accretion disc and $\Omega_{\rm K}$ is the Keplerian angular velocity. {In the gas-pressure-dominated region, we have $\gamma = 1$.}

	In the accretion disc, the Reynolds number ${\rm Re}$ is usually much larger
	than 1, and the hydrodynamic drag force on the magnetic tube follows the
	quadratic law of velocity. For a given density difference on the magnetic tube,
	the drag force per unit tube length is \citep[e.g.,][]{1975ApJ...198..205P, 1979ApJ...230..914P, 1984ApJ...277..312S}
	\begin{equation}
		F_{\rm drag}= \frac{1}{2}\rho_{\rm d}v_{\rm rise}^{2}C_{\rm D}2a, \label{eq:drag}
	\end{equation}
	where $C_{\rm D}$ is the drag coefficient and $v_{\rm rise}$ is the rising
	speed of the magnetic tube. $C_{\rm D}$ is about 1 in the geometry of a cylinder
	and weakly depends on ${\rm Re}$ \citep[e.g.,][]{wieselsberger1922new, goldstein1938modern}.

	The bending of the magnetic tube will generate magnetic tension force which limits
	the upward motion of the magnetic tube. However, the Parker instability requires
	a perturbation wavelength exceeding a critical threshold,
	$\lambda > \lambda_{\rm p, crit}$ \citep[e.g.,][]{1966ApJ...145..811P}. For magnetic
	tubes emerging from the accretion disc, the critical wavelength is usually
	much larger than the scaled height of the accretion disc $H_{\rm d}$. A precise
	analysis of the Parker instability in the nonuniform gravitational field was
	conducted by \citet{1993ApJ...404..185G}, who estimated the minimum critical
	wavelength to be $\lambda_{\rm p, crit}= 2\sqrt{2}\pi H_{\rm d}\approx 8.9H_{\rm
	d}$. The distance between the footpoints of the magnetic tube can be
	estimated by the critical wavelength of Parker instability:
	\begin{equation}
		L \approx{\lambda_{\rm P}},
	\end{equation}
	where $\lambda_{\rm P}$ is the wavelength of Parker instability. For
	magnetic tubes with a wavelength larger than $\lambda_{\rm p, crit}$, the curvature
	radius is large and the tension term is subdominant, thus the terminal
	velocity of the magnetic tube is determined by the balance of buoyancy and
	drag forces.

	Equating (\ref{eq:buoy}) to (\ref{eq:drag}), we obtain the terminal rising velocity:
	\begin{equation}
		v_{\rm rise}= \sqrt{\frac{\pi}{4 \gamma C_{\rm D}}}\sqrt{\frac{\lambda_{\rm
		MRI}}{ H_{\rm d}}}v_{\rm A},
	\end{equation}
	which has a similar form to that derived by \citet{1984ApJ...277..312S}.

	The linear analysis of MRI shows that the MRI will not occur when the typical
	wavelength $\lambda_{\rm MRI}$ is larger than $H_{\rm d}$ \citep[e.g.,][]{BH91},
	which corresponds to the case of extremely strong initial seed magnetic
	fields. Furthermore, $4 C_{\rm D}>\pi$ always holds for cylindrical flux
	tubes at high Reynolds numbers $10<{\rm Re}\lesssim 10^{5}$\citep[e.g.,][]{wieselsberger1922new}.
	This implies that in the standard buoyancy description, the terminal velocity
	of magnetic flux tubes is fundamentally bounded by the local Alfvén speed, i.e.,
	$v_{\rm rise}\lesssim v_{\rm A}$.

	To account for the fact that magnetic emergence and dissipation are
	spatially intermittent, we introduce a surface covering factor $C_{\rm cover}$,
	which is defined as the fraction of the accretion disc surface that is
	magnetically active \citep[e.g.,][]{1994ApJ...432L..95H, 1995ApJ...449L..13S, 1998MNRAS.299L..15D}.
	For highly active regions, $C_{\rm cover}\sim 1$, while for regions where magnetic
	tubes cannot effectively emerge, $C_{\rm cover}\sim 0$. The final expression
	for the magnetic energy transport efficiency in the accretion disc is then given
	by:
	\begin{equation}
		Q_{\rm buoy}= C_{\rm cover}U_{\rm mag}v_{\rm rise}= C_{\rm cover}C_{\rm
		mag}\sqrt{\frac{\pi}{4 \gamma C_{\rm D}}}\sqrt{\frac{\lambda_{\rm MRI}}{H_{\rm
		d}}}\tau_{r\phi}v_{\rm A}, \label{eq:Q_buoy}
	\end{equation}
	For direct comparison with MF02, we also introduce $v_{\rm A}^{({\rm MF02})}$,
	which slightly differs from the standard Alfvén speed $v_{\rm A}$ by a
	factor of $\sqrt{2}$, i.e.
	\begin{equation}
		v_{\rm A}^{({\rm MF02})}= \sqrt{\frac{P_{\rm mag}}{\rho_{\rm d}}}= \frac{
		v_{\rm A}}{\sqrt{2}}.
	\end{equation}
	(\ref{eq:Q_buoy}) can be written as MF02 used form in their paper, i.e.
	\begin{equation}
		Q_{\rm buoy}= b \tau_{r\phi}v_{\rm A}^{({\rm MF02})}, \label{eq:Q_buoy_2}
	\end{equation}
	where $b$ is the efficiency factor. In our paper, $b$ is defined as
	\begin{equation}
		b= C_{\rm cover}C_{\rm mag}\sqrt{\frac{\pi}{2\gamma C_{\rm D}}}\sqrt{\frac{\lambda_{\rm
		MRI}}{H_{\rm d}}}= \frac{b_{0}}{\sqrt{\gamma}}. \label{eq:b}
	\end{equation}

	In the MF02/C09 disc-corona framework, the coronal energy fraction $f$ is
	defined as the ratio between the buoyantly transported (and ultimately dissipated)
	magnetic power and the local viscous dissipation in the thin disc, i.e.
	\begin{equation}
		f= \frac{Q_{\rm buoy}}{Q_{\rm visc}}, \label{eq:f}
	\end{equation}
	where $Q_{\rm visc}$ is the viscous dissipation in the accretion disc. For a
	nearly Keplerian thin disc, the viscous dissipation can be calculated as (e.g.,
	SS73)
	\begin{equation}
		Q_{\rm visc}= \frac{3}{2}\tau_{r\phi}c_{\rm s}, \label{eq:Q_visc}
	\end{equation}
	Thus we can write the energy fraction as
	\begin{equation}
		f = \frac{2}{3}b_{0}\sqrt{\frac{C_{\rm mag}}{\gamma}}\sqrt{\frac{\tau_{r\phi}}{P_{\rm
		tot}}}.
	\end{equation}

	C09 considered three different $\alpha$ prescriptions. The different
	$\alpha$ prescriptions yield different $f-\dot{m}$ relations, i.e.
	\begin{equation}
		f =\frac{2}{3}b_{0}\sqrt{\frac{C_{\rm mag}}{\gamma}}\sqrt{\alpha}\times\left
		\{
		\begin{aligned}
			 & 1,                                                  & \text{for} & ~\tau_{r\phi}= \alpha P_{\rm tot}                   \\
			 & \left(\frac{P_{\rm gas}}{P_{\rm tot}}\right)^{1/4}, & \text{for} & ~\tau_{r\phi}= \alpha \sqrt{P_{\rm tot}P_{\rm gas}} \\
			 & \left(\frac{P_{\rm gas}}{P_{\rm tot}}\right)^{1/2}, & \text{for} & ~\tau_{r\phi}= \alpha P_{\rm gas}
		\end{aligned}\right..
	\end{equation}
	The three prescriptions behave similarly in gas-pressure-dominated region, whereas
	in radiation-pressure-dominated region the two modified prescriptions strongly
	suppress $f$, yielding a trend in which $\langle f\rangle$ decreases as the
	radiation-pressure-dominated region expands with increasing $\dot{m}$.

	However, many observations indicate that the hard X-ray emitting corona is often
	compact and concentrated within a few tens of $R_{\rm g}$. If $f$ is
	systematically suppressed in the radiation-pressure-dominated region, the model
	tends to shift coronal power outward into gas-pressure-dominated regions,
	creating tension with the observed radial compactness of coronal dissipation.

	\subsection{Enhancement by Photon Bubble Instability}
	\label{sec:pbi}

	{Building upon the standard framework, we now introduce PBI into to the model to enhance the vertical energy transport.}

	In the optically thick accretion disc, the radiative transfer equation is well-approximated
	by the diffusion equation \citep[e.g.,][]{1986rpa..book.....R}:
	\begin{equation}
		\mathbf{F}_{\rm rad}= -\frac{c}{3\kappa \rho_{\rm d}}\nabla E,
	\end{equation}
	where $\mathbf{F}_{\rm rad}$ denotes the radiative flux, $\kappa$ represents
	the opacity (dominated by electron scattering in the inner disc), and $E$ is
	the radiation energy density. This relation indicates that within an optically
	thick medium, the radiative diffusion coefficient is inversely proportional
	to the gas density. Consequently, the radiative flux tends to preferentially
	channel through regions of local density deficit \citep[e.g.,][]{1992ApJ...388..561A, 1998MNRAS.297..929G}

	In the radiation-pressure-dominated region, considering both
	$\tau_{r\phi}= \alpha P_{\rm tot}$ and
	$\tau_{r\phi}= \alpha \sqrt{P_{\rm tot}P_{\rm gas}}$ prescription, magnetic pressure
	can be larger than the gas pressure as discussed above. In such a strong
	magnetic field regime, the motion of ionized gas perpendicular to the magnetic
	field lines is strictly constrained. This anisotropic constraint allows
	radiative diffusion to effectively destabilize the slow magnetosonic wave
	and exciting PBI \citep[e.g.,][]{1992ApJ...388..561A,1998MNRAS.297..929G}. This
	mechanism amplifies local density perturbations, eventually evolving into a
	multiphase structure characterized by the coexistence of low-density photon
	bubbles and high-density shocked clumps \citep[e.g.,][]{2001ApJ...551..897B}.

	In the WKB approximation, the dispersion relation for PBI takes a concise form
	\citep[e.g.,][]{1998MNRAS.297..929G,2003ApJ...596..509B}:
	\begin{equation}
		\omega^{2}= -ikg_{\rm d}Q(\hat{k},\hat{b}), \label{eq:pbi_dispersion}
	\end{equation}
	where $k = 2\pi /\lambda_{\rm PBI}$ is the wavenumber, and the geometric factor
	$Q$ depends on the direction of the magnetic field and the wave propagation:
	\begin{equation}
		Q(\hat{k}, \hat{b}) = (\hat{k}\cdot \hat{b}) [\hat{b}\cdot \hat{z}- (\hat
		{k}\cdot \hat{z}) (\hat{k}\cdot \hat{b})].
	\end{equation}
	Here, $\hat{k}$, $\hat{b}$, and $\hat{z}$ denote the unit vectors for the wave
	propagation, the magnetic field, and the vertical direction, respectively. {While MRI turbulence predominantly generates an azimuthal magnetic field in the disc midplane, PBI can still grow for perturbations with non-horizontal wavevectors \citep[e.g.,][]{2003ApJ...596..509B}.}
	Numerical maximization of the $Q$ factor indicates that for a wide range of magnetic
	field directions, the maximum geometric factor is robustly confined to the
	range $0.38\lesssim Q\lesssim 0. 5$. {Once the Parker instability is excited, the buoyant horizontal flux tubes buckle and inevitably develop a non-zero vertical magnetic field component. Even if this vertical component is weak compared to the dominant azimuthal field, it establishes a macroscopic pathway along the field lines, allowing the photon bubbles to be potentially transported to the top of the rising flux tubes.}

	According to (\ref{eq:pbi_dispersion}), the PBI growth rate scales as $\Gamma
	_{\rm PBI}\propto \sqrt{k}$. However, as demonstrated by the rigorous analysis
	of \citet{2003ApJ...596..509B}, the PBI transforms into overstable slow magnetosonic
	modes when $\lambda_{\rm PBI}$ drops below the gas pressure scale height $H_{\rm
	gas}= c_{\rm g}/\Omega_{\rm K}$. In this regime, the growth rate ceases to increase
	with wavenumber and saturates near the scale $\lambda_{\rm PBI}\approx H_{\rm
	gas}$. Setting the characteristic wavenumber $k \approx 2\pi / H_{\rm gas}$,
	the maximum growth rate can be estimated as:
	\begin{equation}
		\Gamma_{\rm PBI}={\rm Im}(\omega) = \sqrt{\frac{kg_{\rm d}Q}{2}}\approx \sqrt{\frac{H_{\rm
		d}}{H_{\rm gas}}}\Omega_{\rm K},
	\end{equation}
	which is even larger than $\Gamma_{\rm MRI}$ in the inner disc.

	By comparison, the typical growth rate of the rising tube is estimated by the
	Alfvén crossing time over the disc scale height:
	\begin{equation}
		\Gamma_{\rm P}\approx\frac{v_{\rm rise}}{H_{\rm d}}\lesssim \frac{v_{\rm
		A}}{H_{\rm d}}.
	\end{equation}
	Consequently, the ratio of the PBI growth rate to the magnetic buoyancy rate
	is:
	\begin{equation}
		\frac{\Gamma_{\rm PBI}}{\Gamma_{\rm P}}\gtrsim \sqrt{\frac{H_{\rm d}}{H_{\rm
		gas}}}\frac{c_{\rm s}}{v_{\rm A}}.
	\end{equation}
	This implies that in the inner disc, the PBI develops and saturates almost instantaneously
	compared to the magnetic buoyancy timescale, justifying the assumption of a
	fully developed porous structure during the flux tube rising.

	In the rising magnetic structures, once PBI enters the nonlinear regime,
	perturbations steepen into propagating shock trains along the magnetic field
	\citep[e.g.,][]{2001ApJ...553..987B}. The flow becomes highly inhomogeneous:
	most of the mass resides in dense shock fronts/sheets, while radiation escapes
	preferentially through low-density gaps, producing a porous medium \citep[e.g.,][]{2003ApJ...596..509B}.
	In the tube, this phase separation naturally reduces the mass loading of the
	rising segment. Dense clumps tend to drain back toward the footpoints under the
	component of gravity along field lines, whereas low-density channels with
	small optical depth $\tau_{\rm in}\approx1$ are maintained by flux
	channeling and expand upward. In the extreme evacuation limit, the density
	contrast approaches $\Delta \rho \approx \rho_{\rm d}$.

	Balancing buoyancy and drag forces, we obtain the characteristic rise velocity
	of the drained flux tube as:
	\begin{equation}
		v_{\rm rise}\approx \sqrt{\frac{\pi}{2C_{\rm D}}\frac{\lambda_{\rm MRI}}{H_{\rm
		d}}}c_{\rm s}. \label{eq:velocity_pbi}
	\end{equation}

	Thus, if PBI efficiently reduces mass loading, rising magnetic structures can
	approach the dynamical, near-sonic limit in the inner disc. We emphasize that
	this high velocity does not violate the fundamental MHD speed limit. The
	relevant limiting Alfvén speed inside the evacuated tube is:
	\begin{equation}
		v_{\rm A,in}= \sqrt{\frac{2P_{\rm mag}}{\rho_{\rm in}}},
	\end{equation}
	which can greatly exceed $c_{\rm s}$ due to the low internal density $\rho_{\rm
	in}\ll \rho_{\rm d}$. Hence, the regime $v_{\rm rise}\sim c_{\rm s}$ is physically
	consistent, even though $v_{\rm rise}$ may significantly exceed the Alfvén
	speed defined using the ambient disc density
	$v_{\rm A}= \sqrt{2P_{\rm mag}/\rho_{\rm d}}$.

	Once the rising speed is determined, we can calculate the magnetic energy transport
	efficiency enhanced by PBI as (\ref{eq:Q_buoy}) and (\ref{eq:Q_buoy_2}):
	\begin{equation}
		Q_{\rm buoy}= b_{\rm PBI}\tau_{r\phi}v_{\rm A}^{({\rm MF02})},
	\end{equation}
	where the efficiency factor $b_{\rm PBI}$ can be larger than $b_{0}$, i.e.
	\begin{equation}
		b_{\rm PBI}= b_{0}\sqrt{\beta}.
	\end{equation}
	According to (\ref{eq:f}) and (\ref{eq:Q_visc}), if the PBI is considered in
	the inner region of the accretion disc, the coronal energy fraction $f$ is
	\begin{equation}
		f = \frac{2}{3}b_{0},
	\end{equation}
	which is not sensitive to $\alpha$.

	To quantify the PBI-enhanced vertical energy transport, we calculate the radial
	distribution of the coronal energy fraction $f$ by comparing the PBI-modified
	model with the standard buoyancy model under different viscosity assumptions
	$\tau_{r\phi}= \alpha P_{\rm tot}$ and $\tau_{r\phi}= \alpha \sqrt{P_{\rm
	tot}P_{\rm gas}}$. We define the following dimensionless physical quantities:
	\begin{equation}
		m = \frac{M_{\rm BH}}{M_{\odot}}, \quad \dot{m}= \frac{\dot{M}}{\dot{M}_{\rm
		Edd}}, \quad r = \frac{R}{R{\rm g}},
	\end{equation}
	where $\dot{M}_{\rm Edd}= L_{\rm Edd}/0.1c^{2}$ is the Eddington accretion rate
	with a radiative efficiency of 0.1. We investigate the $f- r$ profiles for
	typical BH mass $m = 10$ (XRBs) and $m = 10^{7}$ (AGNs) with a moderate
	accretion rate $\dot{m}=0.1$. In our calculations in this paper, we adopt
	fiducial parameters $\alpha = 0.1$ and $b_{0}= 1$. For simplicity, the
	outflow is ignored in the calculation here. We merely demonstrate the enhancing
	energy transport in the inner region with PBI considered. The complete disc
	model with outflow will be established in the next section.

	In practice, the quantity controlling the coronal power fraction is the efficiency
	factor $b$ and we therefore implement the transition between the gas-pressure-dominated
	branch and the radiation-pressure-dominated branch by interpolating $b$, i.e.
	\begin{equation}
		b_{\rm eff}= \frac{P_{\rm gas}^{2}b_{\rm gas}}{P_{\rm gas}^{2}+ P_{\rm
		rad}^{2}}+ \frac{P_{\rm rad}^{2}b_{\rm rad}}{P_{\rm gas}^{2}+ P_{\rm rad}^{2}}
		.
	\end{equation}
	We have verified that using a linear weighting does not change our qualitative
	trends, but it produces a broader transition region around
	$P_{\rm gas}\simeq P_{\rm rad}$.

\begin{figure}
		\includegraphics[width=\columnwidth]{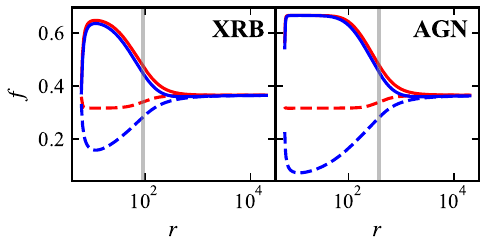}
		\caption{Radial distribution of the coronal energy fraction $f$ for
		different viscosity and magnetic buoyancy prescriptions. The BH mass $m$
		in the left and right panels are 10 and $10^{7}$, respectively. The dimensionless
		accretion rate is $\dot{m}= 0.1$. The red lines represent the standard
		$\alpha$ prescription $\tau_{r\phi}= \alpha P_{\rm tot}$, while the blue
		lines represent the modified $\alpha$ prescription $\tau_{r\phi}= \alpha
		\sqrt{P_{\rm gas}P_{\rm tot}}$. The solid lines represent the situation
		where PBI is taken into account in the radiation-pressure-dominated
		region, while the dashed lines do not consider PBI. The vertical gray
		line indicates the position where the $P_{\rm gas}=P_{\rm rad}$.}
		\label{fig:f-R}
	\end{figure}

	In Fig \ref{fig:f-R}, we evaluate $f(r)$ for different $\alpha$-stress closures
	and illustrate how PBI evacuation reshapes the radial distribution of coronal
	power. It is shown that under the standard $\alpha$ prescription, the coronal
	fraction $f$ remains nearly stable at all positions. Under the modified
	$\alpha$ prescription, the corona fraction $f$ in the inner region is suppressed.
	When considering PBI, the results given by the two $\alpha$ prescriptions are
	similar in the inner region, and $f$ is approximately $\sqrt{\beta}$ larger
	than that in the outer region.

	\section{Global Disc Solutions with MRI-driven Wind}
	\label{sec:global_disc}

	\subsection{Disc Dynamics with Outflows}

	In this section, we construct a steady, axisymmetric thin disc model allowing
	for mass loss into a disc outflow. Mass conservation implies that the
	accretion rate decreases inwards according to
	\begin{equation}
		\frac{{\rm d}\dot{M}}{{\rm d}R}= 4\pi R \,\dot{m}_{\rm w}, \label{eq:mdot_wind}
	\end{equation}
	where $\dot{m}_{\rm w}$ is the outflow mass flux per unit surface area of
	the disc and satisfies
	\begin{equation}
		\dot{m}_{\rm w}= \rho_{0}v_{0}, \label{eq:mdotw}
	\end{equation}
	where $\rho_{0}$ and $v_{0}$ is the initial density and speed of the outflow
	at the disc surface.

	In the presence of outflow, the angular momentum equation can be expressed as
	\begin{equation}
		\frac{{\rm d}}{{\rm d}R}\left(\dot{M}l - 2\pi R^{2}W_{r\phi}\right) = \frac{{\rm d}\dot{M}}{{\rm d}R}
		\,l_{\rm w}, \label{eq:AM_conservative}
	\end{equation}
	where $l= \Omega R^{2}$ is the specific angular momentum of the disc,
	$W_{r\phi}= 2H_{\rm d}\tau_{r\phi}$ is the vertically integrated viscosity
	stress, and $l_{\rm w}$ is the specific angular momentum carried away by the
	outflow at launch. Expanding equation~(\ref{eq:AM_conservative}) gives
	\begin{equation}
		\dot{M}\frac{{\rm d}(\Omega R^{2})}{{\rm d}R}= \frac{{\rm d}}{{\rm d}R}\left
		(2\pi R^{2}W_{r\phi}\right) + \frac{{\rm d}\dot{M}}{{\rm d}R}\left(l_{\rm
		w}-\Omega R^{2}\right). \label{eq:AM_expanded}
	\end{equation}

	In this paper, the accretion disc is nearly Keplerian, so we set $\Omega\approx
	\Omega_{\rm K}$.
	{In the weakly magnetized accretion disc, the wind only serves to reduce the local accretion rate, without extracting additional angular momentum from the accretion disc \citep[e.g.,][]{2018ApJ...857...34Z, 2021A&A...647A.192J}}.
	(\ref{eq:AM_expanded}) can be simplified to
	\begin{equation}
		\frac{1}{2}\dot{M}\Omega_{\rm K}R = \frac{{\rm d}}{{\rm d}R}\left(2\pi R^{2}
		W_{r\phi}\right), \label{eq:AM_simplified}
	\end{equation}
	where $\Omega_{\rm K}=\sqrt{GM_{\rm BH}/R^{3}}$. Integrating (\ref{eq:AM_simplified})
	from the inner boundary $R_{\rm in}$ yields
	\begin{equation}
		W_{r\phi}(R)= \frac{\dot{M}(R)\Omega_{\rm K}(R)}{2\pi}\,\mathcal{F}(R), \label{eq:W_rphi}
	\end{equation}
	where $\mathcal{F}$ is the boundary-condition correction factor. For a zero-torque
	inner boundary, $\mathcal{F}$ can be written as
	\begin{equation}
		\mathcal{F}(R)= \frac{\int_{R_{\rm in}}^{R}\dot{M}(R')\,R'^{-1/2}\,{\rm d}R'}{2\,\dot{M}(R)\,R^{1/2}}
		. \label{eq:Ffactor}
	\end{equation}
	For a constant $\dot{M}(R)$, this reduces to the standard thin disc result $\mathcal{F}
	=1-(R_{\rm in}/R)^{1/2}$. In our calculations, the inner boundary is set at the
	innermost stable circular orbit (ISCO) of the black hole, i.e.,
	$r_{\rm in}= 6$.

	The viscous dissipation rate per unit surface area of the disc is
	\begin{equation}
		Q_{\rm visc}= -\frac{1}{2}W_{r\phi}\left(R\frac{{\rm d}\Omega_{\rm K}}{{\rm d}R}
		\right) = \frac{3}{4}\Omega_{\rm K}W_{r\phi},
	\end{equation}
	and using (\ref{eq:W_rphi}) we obtain
	\begin{equation}
		Q_{\rm visc}= \frac{3}{8\pi}\frac{GM_{\rm BH}\dot{M}(R)}{R^{3}}\,\mathcal{F}
		(R). \label{eq:Qvisc_wind}
	\end{equation}
	In practice, $\mathcal{F}(R)$ is evaluated by numerically integrating (\ref{eq:Ffactor})
	outward from $R_{\rm in}$ once $\dot{M}(R)$ is specified by a wind prescription.

	In our model, the magnetic energy transported from the disc into the corona
	by buoyancy is not only available for heating the coronal plasma, but also consumed
	in mass loading and accelerating a slow outflow. For simplicity, we assume that
	a fraction of the local viscous dissipation $fQ_{\rm visc}$ is transported into
	the corona, while the remaining fraction $(1-f)Q_{\rm visc}$ is radiated
	locally by the disc. The coronal irradiation and reprocessing of the disc is
	neglected in the present work \citep[e.g.,][also see MF02]{1994ApJ...436..599S}.

	Under the diffusion approximation for an optically thick disc, the radiative
	flux satisfies
	\begin{equation}
		(1-f)Q_{\rm visc}= \frac{2acT_{\rm d}^{4}}{3\kappa\Sigma}= \sigma T_{\rm
		eff}^{4}, \label{eq:Qrad_disk}
	\end{equation}
	where $T_{\rm d}$ is the mid-plane temperature, $T_{\rm eff}$ is the
	effective temperature at the disc surface, $\Sigma=2\rho_{\rm d}H_{\rm d}$ is
	the surface density, $\kappa = \kappa_{\rm ff}+ \kappa_{\rm es}$ is the mean
	opacity, and $a=4\sigma/c$ is the radiation constant.

	\citet{2013MNRAS.430.1102T} (hereafter T13) performed a joint analysis of
	UFOs and warm absorbers (WAs) in a sample of Seyfert 1 galaxies, and found that
	the outflow properties vary continuously with the inferred distance from the
	black hole, supporting a stratified disc-wind picture. In their paper, they
	reported the distribution of
	$\dot{m}_{\rm out}=\dot{M}_{\rm out}/\dot{M}_{\rm Edd}$ as a function of
	dimensionless radius $r$ (see their Fig. 3d), where $\dot{M}_{\rm out}$ is
	the total mass outflow rate estimated for each outflow component and $r$ is
	the characteristic location inferred from the photoionization and escape-velocity
	constraints.

	For a qualitative and scale-free comparison with our theoretical upper
	limits, we describe the empirical $\dot{m}_{\rm out}$-$r$ trend using a simple
	power-law accretion-rate profile,
	\begin{equation}
		\dot{m}(r) = \dot{m}_{\rm in}\left(\frac{r}{r_{\rm in}}\right)^{s}, \label{eq:mdot_powerlaw}
	\end{equation}
	so that the differential mass loading per logarithmic radius,
	\begin{equation}
		\frac{{\rm d} \dot{M}}{{\rm d}\ln R}= 4\pi R^{2}\dot{m}_{\rm w}(R) = s~\dot
		{M}_{\rm in}\left(\frac{r}{r_{\rm in}}\right)^{s}. \label{eq:dMw_dlnR}
	\end{equation}
	We calculate the predicted cumulative outflow rate as $\dot{m}_{\rm out}(r) =
	\dot{m}(r)-\dot{m}_{\rm in}$ and fit the combined UFO+WA dataset in T13 with
	(\ref{eq:mdot_powerlaw}) and obtain the best-fitting parameters $\dot{m}_{\rm
	in}= 0.10\pm 0.07$ and $s= 0.14 \pm 0.05$.

	To assess whether the coronal power budget is sufficient to drive the empirically
	inferred mass loading, we compute the maximum wind mass flux
	$\dot{m}_{\rm w, max}$ allowed by the local coronal dissipation. We define $\dot
	{m}_{\rm w, max}$ as a conservative energetic upper limit obtained by assuming
	that the entire coronal power per unit surface area $Q_{\rm buoy}$ is used only
	to overcome the gravitational binding energy of gas on a Keplerian orbit:
	\begin{equation}
		\dot{m}_{\rm w, max}(R) = \frac{f Q_{\rm visc}}{GM_{\rm BH}/2R}. \label{eq:mwmax}
	\end{equation}

\begin{figure}
		\includegraphics[width=\columnwidth]{
			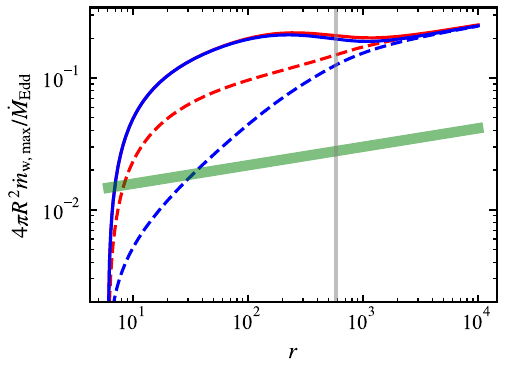
		}
		\caption{Radial distribution of the inferred upper limit on wind mass
		loading for different viscosity and magnetic buoyancy prescriptions. The
		BH mass is $m_{\rm BH}=10^{7}$. The meanings of different line styles (red/blue
		and solid/dashed) are the same as those in the Fig. \ref{fig:f-R}. The
		bold green line represents the empirical mass-loading profile of the
		wind implied by the combined UFO+WA sample of T13. The vertical gray line
		indicates the position where the $P_{\rm gas}=P_{\rm rad}$.}
		\label{fig:mdotw-R}
	\end{figure}

	In the Fig. \ref{fig:mdotw-R}, we compares $4\pi R^{2}\dot{m}_{\rm w, max}/\dot
	{M}_{\rm Edd}$ predicted by different prescriptions (different $\alpha$
	prescriptions and with/without PBI mechanism) to the empirical $4\pi R^{2}\dot
	{m}_{\rm w}$ profile implied by the fitting. We find that, in the modified-$\alpha$
	case without PBI (i.e. MF02 model), the coronal power is insufficient to support
	the inferred outflow at $r\lesssim 40$, as the upper limit falls below the
	empirical requirement. Such deviations are difficult to be compensated for by
	adopting a larger $\alpha$ in the calculation, because $f$ is only
	proportional to $\sqrt{\alpha}$ in the MF02 model. In contrast, for the
	standard-$\alpha$ case without PBI, the inferred mass loading can be maintained
	down to $r\approx 10$. However, once PBI-enhanced buoyant transport is included,
	both $\alpha$ prescriptions yield very similar results: the upper-limit curves
	remain above the empirical requirement over essentially the full radial range
	and still provide substantial headroom at $r\sim 10$, indicating that only a
	modest fraction of the coronal power would be needed to account for the
	empirical mass loading, while the remaining power can be dissipated to accelerate
	the outflow or heat the corona. This is consistent with recent research that
	the minimum UFO launching radius may approach the ISCO \citep[e.g.][]{2025arXiv251206077L}.

	In Sec. \ref{sec:pbi}, we have shown that under the standard and modified
	$\alpha$ prescription, the coronal fraction $f$ differs little when PBI is considered.
	Therefore, to avoid the complexity of the discussion, we adopted the
	standard $\alpha$ prescription to construct the model, which is widely used in
	current accretion disc researches. The PBI-induced evacuation is also
	considered in the calculations that follow.

	\subsection{Mass Loading Prescription and Observational Constraints}

	To construct a disc-corona model with mass loss, a prescription for the wind
	mass loading is required. A detailed discussion of possible wind-driving mechanisms
	and their limitations will be presented in Paper II. Here we adopt a
	physically motivated fiducial closure in which the outflow is parameterised
	by a vertical torque.

	If a vertical torque $\Lambda_{z\phi}$ acts on per unit surface area of the
	disc and extracts angular momentum from the disc into the outflow, then the mass
	flux at the outflow base can be written as
	\begin{equation}
		\dot{m}_{\rm w}= \frac{\Lambda_{z\phi}}{l_{\rm w}-\Omega_{\rm K}R^{2}}= \frac{\Lambda_{z\phi}}{(\lambda
		- 1)\Omega_{\rm K}R^{2}}, \label{eq:mdot_wind_torque}
	\end{equation}
	where the laver arm parameter $\lambda = l_{\rm w}/\Omega_{\rm K}R^{2}$ measures
	the specific angular momentum carried by the outflow relative to the local Keplerian
	value of the accretion disc.

	In the torque-based wind models, an $\alpha$-type prescription is commonly
	used to parameterize the vertical torque \citep[e.g.,][]{2022MNRAS.512.2290T, 2025PTEP.2025b3E02T},
	i.e.
	\begin{equation}
		\Lambda_{z\phi}= \alpha_{z\phi}P_{\rm tot}H_{\rm d}. \label{eq:Lambda_zphi}
	\end{equation}
	Then (\ref{eq:mdot_wind_torque}) can be written as
	\begin{equation}
		\dot{m}_{\rm w}= \frac{\alpha_{z\phi}}{\lambda-1}\left(\frac{H_{\rm d}}{R}
		\right)^{2}\rho_{\rm d}c_{\rm s}. \label{eq:mdot_wind_alpha}
	\end{equation}

	{A large-scale magnetocentrifugal wind can in principle enforce a large lever arm ($\lambda\gg 1$) and efficiently extract angular momentum,}
	but it typically requires a sufficiently strong net poloidal magnetic flux and
	a coherent field geometry, which may be difficult to realise in thin discs. Consistently,
	global MHD simulations suggest that magnetocentrifugal winds are readily
	produced only when the imposed vertical field is strong (e.g.
	$\beta_{0}\lesssim 10^{3}$), whereas in the weak-field regime ($\beta_{0}> 10
	^{3}$) the flow is dominated by MRI-driven turbulence \citep[e.g.,][]{2013ApJ...767...30B}.
	Here $\beta_{0}$ represents the initial plasma $\beta$ value set in the MRI numerical
	simulation.

	However, even in weak-field discs, MRI turbulence can generate Alfv\'en and magnetosonic
	waves that propagate toward the disc surface. As the density drops with
	height, these waves can steepen into nonlinear shocks and transfer a fraction
	of the turbulent Poynting flux into gas kinetic energy, thereby loading mass
	into the outflow \citep[e.g.,][]{2009ApJ...691L..49S, 2010ApJ...718.1289S, 2013ApJ...767...30B}.
	We refer to this turbulence-driven mass-loading channel as the Suzuki-Inutsuka
	(SI) mechanism.
	{In contrast to magnetocentrifugal winds, in recent MHD simulations, turbulence-driven outflows are found to develop in the weak-field regime even with $\beta_{0}>10^{4}$ and are expected to extract little angular momentum \citep[e.g.,][]{2018ApJ...857...34Z, 2019MNRAS.490.3112J, 2021A&A...647A.192J}.}
	\footnote{We note that this global weak-field condition ($P_{\rm mag}< P_{\rm
	tot}$) is perfectly compatible with the local strong-field requirement for
	PBI ($P_{\rm mag}\gtrsim P_{\rm gas}$) in the radiation-pressure-dominated
	inner regime. The magnetic pressure can easily exceed the local gas pressure
	in the inner region to trigger PBI, while still remaining a small fraction
	of the total pressure ($P_{\rm mag}\sim \alpha P_{\rm tot}< P_{\rm tot}$). This
	ensures the global magnetic field is too weak to extract significant angular
	momentum.}

	{In semi-analytical frameworks, the lever arm parameter $\lambda$ depends on the radial wind profile, i.e. ${\rm d}\ln \dot{M}/{\rm d}\ln R$, and the ratio between the vertical and radial torque, i.e. $\alpha_{z\phi}/ \alpha$. Under the conditions considered in this paper, $\lambda$ is expected to be moderately larger than unity \citep[e.g.,][]{2019MNRAS.490.3112J, 2022MNRAS.512.2290T}. Given the uncertainties in these exact values, we absorb the uncertain combination $\alpha_{z\phi}/\lambda -1$ into an effective wind efficiency parameter,}
	\begin{equation}
		\alpha_{\rm w}=\frac{\alpha_{z\phi}}{\lambda-1},
	\end{equation}
	with which (\ref{eq:mdot_wind_alpha}) can be written as
	\begin{equation}
		\dot{m}_{\rm w}= \alpha_{\rm w}\left(\frac{H_{\rm d}}{R}\right)^{2}\rho_{\rm
		d}c_{\rm s}. \label{eq:SIwind}
	\end{equation}
	While the velocity amplitude of the waves reaches the order of the sound speed
	to trigger nonlinear shocks, the net vertical mass flux is carried by a bulk
	flow that is significantly subsonic in the launch region. In our calculations,
	we set $v_{0}= 0.1 c_{\rm s, surf}$ as the initial wind velocity, consistent
	with the characteristic values seen in numerical experiments \citep[e.g.,][]{2009ApJ...691L..49S,2018ApJ...857...34Z}.

	Under the conditions considered in this work, $\lambda$ is expected to be of
	order unity, rather than exactly 1

	\begin{figure}
		\includegraphics[width=\columnwidth]{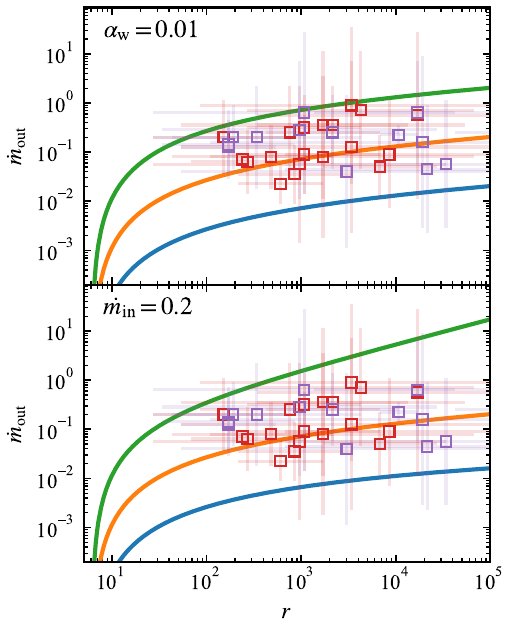}
		\caption{Radial distribution of the cumulative outflow rate for
		different values of $\alpha_{\rm w}$ and $\dot{m}_{\rm in}$. The standard
		$\alpha$ prescription and PBI mechanism are included. In the upper panel,
		$\alpha_{\rm w}$ is fixed to $0.01$, from bottom to top the curves
		correspond to $\dot{m}_{\rm in}=0.02, 0.2$ and $2$, respectively. In the
		lower panel, $\dot{m}_{\rm in}$ is fixed to $0.2$, from bottom to top
		the curves correspond to $\alpha_{\rm w}=0.001, 0.01$ and $0.1$, respectively.
		The red and purple points are the UFO samples of T13 and G15, respectively.}
		\label{fig:SI-mdotw}
	\end{figure}

	Fig. \ref{fig:SI-mdotw} provides a calibration check for the SI mechanism
	adopted in this paper. We compare the predicted cumulative outflow rate against
	the UFO samples of T13 and \citet{2015MNRAS.451.4169G} (hereafter G15), which
	are mainly located within the region of $r\lesssim 10^{4}$. The samples in
	the G15 are processed in the same way as T13 to get the radial profiles of
	physical quantities. We find that a fiducial choice $\dot{m}_{\rm in}\approx
	0.2$ and $\alpha_{\rm w}\approx 0. 01$ is roughly in agreement with the observed
	locus of the UFO samples. The fiducial value of $\alpha_{\rm w}$ is also in
	agreement with the result obtained in the numerical simulation under the
	weak field limit \citep[e.g.,][]{2009ApJ...691L..49S, 2010ApJ...718.1289S}.

	\section{The Simplified Outflow-based Corona Model}
	\label{sec:model}

	The corona suspended above a thin accretion disc is commonly inferred to be magnetically
	dominated (i.e. $\beta \ll 1$), where gas dynamics are primarily controlled by
	the magnetic field. Upon emerging from the disc, magnetic flux tubes allow the
	internal gas to drain downwards under gravity, leaving the rising tubes in a
	highly evacuated situation. As will be demonstrated in Paper II, for an
	evacuated flux tube rising into a background magnetic field that decays with
	height, the maximum ascent height under the force-free approximation is determined
	by the footpoint separation $L$ and the specific decay profile of the background
	field. We find that this maximum ascent height is roughly comparable to $L$
	across various common background magnetic field profiles.

	Consequently, in the absence of violent eruptive events, the vertical extent
	of the corona is essentially set by the characteristic wavelength of the
	Parker instability. \citet{1991IAUS..144..429M} showed that in a high-$\beta$
	disc environment, the maximum wavelength of the Parker instability can reach
	approximately $\lambda_{\rm p, max}\simeq (3.5 \beta + 6) H_{\rm d}$. For
	the disc environment in this paper, this yields $\lambda_{\rm p, max}\approx
	20 H_{\rm d}$. Therefore, the maximum height to which magnetic buoyancy can transport
	energy in a quasi-steady state is $H_{\rm max}\sim 20 H_{\rm d}$, which can
	be taken as a canonical coronal thickness $H_{\rm c}$.

	In Section \ref{sec:global_disc}, we discussed the mass loading of the MRI driven
	outflow from the disc surface. However, the initial velocity of the wind driven
	by the weakly magnetized disc surface is typically low, which is far
	insufficient to escape the gravitational potential well of the black hole. Further
	acceleration within the corona is therefore required.

	As the gas enters the corona from the base, it experiences a vertical
	magnetic pressure gradient exerted by the decaying magnetic field. This force
	works against gravity and accelerates the flow \citep[e.g.,][]{1996MNRAS.279..389L, 2003MNRAS.341.1360L}.
	The vertical hydrodynamic equation for the fluid can be written as:
	\begin{equation}
		\rho v_{z}\frac{\partial v_{z}}{\partial z}= -\frac{\partial P_{\rm gas,
		c}}{\partial z}- \frac{\partial P_{\rm mag, c}}{\partial z}- \rho g_{z},
		\label{eq:MHD_momentum}
	\end{equation}
	where $P_{\rm gas, c}$ and $P_{\rm mag, c}$ denote the local gas and magnetic
	pressures within the corona, respectively. These two gradient terms represent
	the thermal and magnetic driving mechanisms.

	Assuming negligible lateral expansion for the vertical streamline, mass
	conservation requires:
	\begin{equation}
		\rho v_{z}= \dot{m}_{\rm w}, \label{eq:MHD_mass}
	\end{equation}
	{where $\dot{m}_{\rm w}$ is the mass-loss rate of the wind, which is computed explicitly from (\ref{eq:SIwind}) and assumed to remains constant with height}.

	In the magnetically dominated corona regime ($\beta \ll 1$), the gas
	pressure gradient is subdominant compared to the magnetic force. Neglecting the
	thermal driving term and multiplying Equation (\ref{eq:MHD_momentum}) by $v_{z}$,
	we obtain the MHD energy equation along the streamline:
	\begin{equation}
		\dot{m}_{\rm w}\frac{\partial}{\partial z}\left(\frac{1}{2}v_{z}^{2}\right
		) = q_{\rm mec}- \dot{m}_{\rm w}g_{z}, \label{eq:MHD_momentum_mag}
	\end{equation}
	where $q_{\rm mec}= -v_{z}~\partial P_{\rm mag, c}/\partial z$ represents the
	mechanical power of the magnetic pressure gradient. In a steady-state corona,
	this consumption is continuously replenished by the Poynting flux injected from
	the disc via magnetic buoyancy. Beyond the coronal boundary ($z > H_{\rm c}$),
	this magnetic replenishment ceases, and the outflow transitions into a
	thermally or inertia driven free wind.

	Without assuming a specific profile for the background magnetic field, we can
	integrate Equation (\ref{eq:MHD_momentum_mag}) vertically across the corona (from
	$H_{\rm d}$ to $H_{\rm c}$) to obtain the global mechanical energy balance:
	\begin{equation}
		Q_{\rm mec}= \dot{m}_{\rm w}\left(\frac{1}{2}v_{z}^{2}+ \Phi\right)\bigg|
		_{H_{\rm d}}^{H_{\rm c}}, \label{eq:terminal_v}
	\end{equation}
	where $\Phi = -GM / \sqrt{R^{2}+ z^{2}}$ is the gravitational potential per
	unit mass and $Q_{\rm mec}$ is the total magnetic mechanical work per unit
	area.

	Given the inherent difficulties in constraining the detailed vertical
	distribution of the magnetic field and the microphysics of magnetic
	reconnection, we adopt a parametric approach to estimate the terminal velocity
	of the accelerated gas. We assume that a fraction $f_{\rm acc}$ of the
	magnetic energy transported into the corona via buoyancy is converted into
	mechanical work by magnetic pressure gradients, such that $Q_{\rm mec}= f_{\rm
	acc}Q_{\rm buoy}$.

\begin{figure}
		\includegraphics[width=\columnwidth]{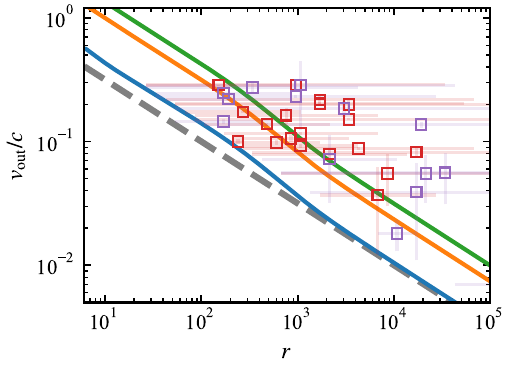}
		\caption{Radial profiles of the outflow velocity at the top of the
		corona derived from our model. The solid coloured curves, from bottom to
		top, correspond to mechanical efficiencies of $f_{\rm acc}= 0.1, 0.5,$ and
		$0.9$, respectively. The dashed line indicates the local Keplerian velocity,
		$v_{\rm K}= \sqrt{GM/R}$. The red and purple points are the UFO samples of
		T13 and G15, respectively.}
		\label{fig:velocity_profile}
	\end{figure}

	In Fig. \ref{fig:velocity_profile}, we present the gas velocity at the top of
	the corona $v_{\rm out}$ calculated for various values of the mechanical efficiency
	parameter $f_{\rm acc}$. Comparing our model predictions with the
	observational data from T13, we find that in the inner disc region ($R \lesssim
	10^{3}R_{\rm g}$), a high mechanical efficiency of $f_{\rm acc}\gtrsim 0.5$
	is required to accelerate the gas to the observed velocities solely via
	coronal MHD processes.

	Conversely, in the gas-pressure-dominated region ($R > 10^{3}R_{\rm g}$),
	even a very high mechanical efficiency $f_{\rm acc}\rightarrow 1$ fails to
	reproduce the high observed velocities. This discrepancy suggests that additional
	acceleration mechanisms may become dominant at larger radii. For instance, as
	the ionization parameter $\xi$ decreases in the outer disc, radiation
	pressure on spectral lines (line driving) could play an increasingly
	significant role in boosting the outflow velocity \citep[e.g.][]{2000ApJ...543..686P}.

	It is important to note that our calculated $v_{\rm out}$ represents the velocity
	at the upper boundary of the coronal heating zone. Beyond this region, where
	magnetic buoyancy ceases to supply energy, the substantial enthalpy of the
	high-temperature coronal gas may drive further acceleration via thermal
	process, potentially bridging the gap between our model predictions and observations.
	Furthermore, we note that even with a modest efficiency of $f_{\rm acc}=0.1$,
	the terminal velocity at the coronal top exceeds the local Keplerian
	velocity, ensuring that gas in the near-Keplerian orbit becomes energetically
	unbound and can escape the system.

	When a fraction $f_{\rm acc}$ of the buoyant magnetic energy flux is
	converted into mechanical work to accelerate the outflow, the remaining energy
	is available for coronal heating via magnetic reconnection. In the limit of
	efficient dissipation, we can express the maximum heating rate as $Q_{\rm
	heat}= (1 - f_{\rm acc})Q_{\rm buoy}$. In the low-$\beta$ environment, magnetic
	reconnection is expected to heat ions preferentially, with only a sub-dominant
	fraction of the energy allocated to electrons. We define $\delta$ as the fraction
	of dissipated energy that directly heats electrons. Experimental and
	observational studies of low-$\beta$ reconnection suggest
	$\delta \approx 1/3$ \citep[e.g.,][]{2013PhRvL.110v5001E, 2014NatCo...5.4774Y}.

	Since we don't know the details of magnetic energy release at different
	heights, we adopt a vertically isothermal approximation for the corona. The
	thermal energy gained by ions from magnetic reconnection is balanced by
	energy transfer to electrons via Coulomb collisions and the advection of
	energy carried away by the outflow. The ion energy equation is thus given by:
	\begin{equation}
		(1 - \delta) Q_{\rm heat}= Q_{\rm ie}+ \frac{\gamma}{\gamma - 1}\frac{\dot{m}_{\rm
		w}}{\mu_{\rm i}m_{\rm p}}k_{\rm B}T_{\rm i},
	\end{equation}
	where $\mu_{\rm i}= 1.23$ and $\gamma = 5/3$ is adopted here. Given the similarity
	in mass, we assume thermal equilibrium between hydrogen and helium ions. The
	Coulomb coupling rate $Q_{\rm ie}$ for a relativistic plasma is \citep[e.g.,][]{1983MNRAS.204.1269S, 1998MNRAS.296L..51Z}:
	\begin{equation}
		\begin{aligned}
			Q_{\rm ie}= & H_{\rm c}\sum_{j}\frac{3}{2}\frac{Z_j^2}{A_j}\frac{m_{\rm e}}{m_{\rm p}}n_{\rm e}n_{j}\sigma_{\rm T}c \frac{(k_{\rm B} T_{\rm e} - k_{\rm B} T_{\rm i})}{K_{2}(1/\theta_{\rm e}) K_{2}(1/\theta_{j})}\ln \Lambda                                                   \\
			            & \times \left[ \frac{2(\theta_{\rm e} + \theta_{j})^{2} + 1}{\theta_{\rm e} + \theta_{j}}K_{1}\left(\frac{\theta_{\rm e} + \theta_{j}}{\theta_{\rm e}\theta_{j}}\right) + 2 K_{0}\left(\frac{\theta_{\rm e} + \theta_{j}}{\theta_{\rm e}\theta_{j}}\right) \right],
		\end{aligned}
	\end{equation}
	where the summation runs over all ionic species, $\theta_{\rm e}= k_{\rm B}T_{\rm
	e}/ m_{\rm e}c^{2}$ and $\theta_{j}= k_{\rm B}T_{\rm i}/ A_{j}m_{\rm p}c^{2}$
	are the dimensionless temperatures for electrons and ions, respectively, and
	$\ln \Lambda$ is the Coulomb logarithm.

	For electrons, the main radiation cooling mechanism in the corona is inverse
	Compton scattering, compared with which bremsstrahlung and synchrotron radiation
	are subdominant (e.g., C09). The electron energy balance equation is
	therefore:
	\begin{equation}
		Q_{\rm ie}+ \delta Q_{\rm heat}= Q_{\rm comp}+ \frac{\gamma}{\gamma - 1}\frac{\dot{m}_{\rm
		w}}{\mu_{\rm e}m_{\rm p}}k_{\rm B}T_{\rm e},
	\end{equation}
	where $\mu_{\rm e}= 1.14$. The Compton cooling rate $Q_{\rm comp}$ can be
	approximated as \citep[e.g.,][]{1986rpa..book.....R, 2013LNP...873.....G}:
	\begin{equation}
		Q_{\rm comp}\approx (A - 1) F_{\rm in}= (e^{y}- 1) \sigma T_{\rm eff}^{4}
		, \label{eq:compton_norm}
	\end{equation}
	where $\sigma T_{\rm eff}^{4}$ is the energy flux of seed photon from the underlying
	accretion disc. The Compton amplification factor is $A = e^{y}$, where the Compton
	$y$-parameter is:
	\begin{equation}
		y = 4\theta_{\rm e}(1 + 4\theta_{\rm e})(\tau_{\rm c}+ \tau_{\rm c}^{2}),
	\end{equation}
	and $\tau_{\rm c}= n_{\rm e}\sigma_{\rm T}H_{\rm c}$ is the vertical Thomson
	optical depth of the corona.

	Since we do not solve the full vertical momentum equation in this work (a detailed
	dynamical treatment is reserved for Paper II), the exact vertical density profile
	remains unconstrained. This introduces an uncertainty in determining the
	effective electron density $n_{\rm e}$ required to evaluate $Q_{\rm comp}$
	and $Q_{\rm ie}$. To close the system of equations, we assume a simple
	velocity profile to estimate the density structure. Specifically, we adopt a
	quadratic velocity law:
	\begin{equation}
		v_{z}(z) = v_{0}+ \frac{v_{\rm out}- v_{0}}{H_{\rm c}^{2}}(z - H_{\rm d})
		^{2}.
	\end{equation}
	From Eq. (\ref{eq:MHD_mass}), this velocity profile implies a density distribution
	roughly proportional to $z^{-2}$, which has a similar form to the conical
	wind with constant velocity and opening angle \citep[e.g.,][]{1995ApJ...447..512K}.
	Assuming $v_{\rm out}\gg v_{0}$, we can derive the effective mean electron
	density $\langle n_{\rm e}\rangle$ and mean squared density
	$\langle n_{\rm e}^{2}\rangle$ by averaging over the vertical extent:
	\begin{equation}
		\langle n_{\rm e}\rangle = \left\langle \frac{\dot{m}_{\rm w}}{\mu_{\rm
		e}m_{\rm p}v_{z}}\right\rangle \approx \frac{\pi}{2}\frac{\dot{m}_{\rm w}}{\mu_{\rm
		e}m_{\rm p}v_{\rm out}}\sqrt{\frac{v_{\rm out}}{v_{0}}},
	\end{equation}
	and
	\begin{equation}
		\langle n_{\rm e}^{2}\rangle = \left\langle \left( \frac{\dot{m}_{\rm w}}{\mu_{\rm
		e}m_{\rm p}v_{z}}\right)^{2}\right\rangle \approx \frac{\pi}{4}\left( \frac{\dot{m}_{\rm
		w}}{\mu_{\rm e}m_{\rm p}v_{\rm out}}\right)^{2}\left( \frac{v_{\rm out}}{v_{0}}
		\right)^{1.5}.
	\end{equation}

	For a corona characterized by a local electron temperature $T_{\rm e}$ and
	vertical optical depth $\tau_{\rm c}$, the Comptonization of soft seed photons
	produces an X-ray spectrum that exhibits a power-law shape,
	$F_{\nu}\propto \nu^{-\alpha_{\rm spec}}$. The spectral index $\alpha_{\rm
	spec}$ can be approximated by the analytical expression \citep[e.g.,][]{1980A&A....86..121S, 1994ApJ...434..570T}:
	\begin{equation}
		\alpha_{\rm spec}= \sqrt{\frac{9}{4} + \gamma}- \frac{3}{2},
	\end{equation}
	where $\gamma$ is a dimensionless parameter, which depends on the local coronal
	temperature $T_{\rm e}$ and optical depth $\tau_{\rm c}$. Here we adopt the
	expression proposed by \citet{1994ApJ...434..570T} in this paper.

\begin{figure}
		\includegraphics[width=\columnwidth]{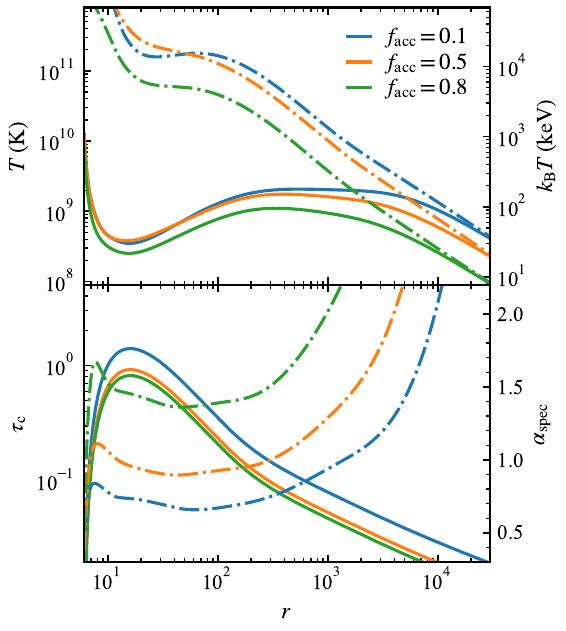}
		\caption{Radial structure of the corona for different mechanical
		efficiencies $f_{\rm acc}$. The upper panel shows the profiles of ion temperature
		$T_{\rm i}$ (dash-dotted curves) and electron temperature $T_{\rm e}$ (solid
		curves) and the scales in K and keV units are all marked in the figure.
		The lower panel shows the profiles of the vertical Thomson optical depth
		$\tau_{\rm c}$ of the corona (solid curves) and the resulting energy
		spectral index $\alpha_{\rm spec}$ (dash-dotted curves).}
		\label{fig:corona_profile}
	\end{figure}

	In Fig. \ref{fig:corona_profile}, we present the radial profiles of the ion and
	electron temperatures, vertical optical depth, and the resulting spectral index
	for different values of the mechanical efficiency parameter $f_{\rm acc}$.
	As $f_{\rm acc}$ increases, a larger fraction of the magnetic energy is diverted
	to drive the high-speed outflow, leaving a reduced budget for coronal
	heating. Consequently, both the ion and electron temperatures decrease
	monotonically with increasing $f_{\rm acc}$. Our model can produce compact corona
	with the optical depth $\tau_{\rm c}$ peaking in the range $10 \lesssim r \lesssim
	100$, where the mass loading is most effective. In the gas-pressure-dominated
	region ($r > 1000$), the corona becomes quite optically thin ($\tau_{\rm c}\ll
	1$), rendering Compton scattering inefficient.

	A notable feature of our model is the sharp decline in electron temperature near
	$r \approx 10$, coincident with the peak in optical depth. This indicates a high
	concentration of coronal material in the inner region, which facilitates
	efficient inverse Compton cooling, thereby suppressing the electron temperature.
	{Although the electron temperature in the outer corona is higher than that in the inner region, our numerical integration reveals that its contribution to the total X-ray emission is negligible because the energy dissipation rate is extremely low. The coronal electron temperature inferred from the observed high-energy cut-off reflects a luminosity-weighted average of the inner coronal regions}

	Regarding the spectral shape, as $f_{\rm acc}$ increases, the spectrum softens
	(i.e., $\alpha_{\rm spec}$ increases). The minimum of the spectral index
	$\alpha_{\rm spec}$ is found in the region $10 < r < 100$, where the bulk of
	the X-ray emission is produced. For moderate acceleration efficiencies, the values
	cluster around $\alpha_{\rm spec}\approx 1$, which is in excellent agreement
	with the characteristic X-ray spectral slopes observed in AGNs \citep[e.g.,][]{1994MNRAS.268..405N, 1996MNRAS.283..193Z}.

	\section{Observational properties of the corona with different accretion
	rate}
	\label{sec:varyingmdot}

	In the previous sections, we have shown that the inner corona can simultaneously
	drive a fast outflow and produce intense radiation, we now investigate
	whether the predicted coronal properties respond to changes in the accretion
	rate in a manner consistent with observations. We perform calculations with varying
	inner accretion rates $\dot{m}_{\rm in}$, while fixing the mechanical
	efficiency at $f_{\rm acc}= 0.5$.

	We first examine the impact of $\dot{m}_{\rm in}$ on the terminal wind
	velocity. Remarkably, our numerical results indicate that the radial profile
	of the terminal velocity remains virtually unchanged within the inner region
	as the accretion rate varies. This insensitivity can be understood through a
	scaling analysis of the radiation-pressure-dominated zone.

	In the standard accretion disc model, $H_{\rm d}$, $\rho_{\rm d}$, and
	$c_{\rm s}$ in (\ref{eq:SIwind}) scale as \citep[e.g.,][]{2002apa..book.....F,2008bhad.book.....K}:
	\begin{equation}
		\begin{aligned}
			 & H_{\rm d}\propto m \dot{m}\mathcal{F},                                    \\
			 & \rho_{\rm d}\propto \alpha^{-1}m^{-1}\dot{m}^{-2}r^{3/2}\mathcal{F}^{-2}, \\
			 & c_{\rm s}\propto \dot{m}r^{-3/2}\mathcal{F}.
		\end{aligned}
	\end{equation}
	Substituting these into our adopted mass-loading prescription, we derive the
	scaling relation for the wind mass flux:
	\begin{equation}
		\dot{m}_{\rm w}\propto \alpha^{-1}m^{-1}\dot{m}r^{-2}\mathcal{F}.
	\end{equation}
	This implies that the mass loading rate scales linearly with the accretion rate,
	$\dot{m}_{\rm w}\propto \dot{m}$. Therefore the total mass loss rate of the outflow
	$\dot{m}_{\rm out}$ is proportional to $\dot{m}_{\rm in}$, which is roughly
	in agreement with the observed results \citep[e.g.,][also see G15]{2017A&A...601A.143F}.

	Since the magnetic energy supply from buoyancy is also proportional to the accretion
	power ($Q_{\rm mag}= fQ_{\rm visc}\propto m^{-1}\dot{m}r^{-3}\mathcal{F}$),
	the energy available per unit mass of the outflow remains insensitive to
	$\dot{m}$. We have shown in Fig. \ref{fig:velocity_profile} that only a
	small fraction of the energy is used to overcome the gravitational potential
	energy in the corona region. Assuming the initial launch velocity $v_{0}$ is
	negligible and the parameters $f$ and $f_{\rm acc}$ do not change with $\dot{m}
	_{\rm in}$, according to (\ref{eq:terminal_v}), the terminal velocity can be
	estimated as:
	\begin{equation}
		v_{\rm out}\approx \sqrt{2f_{\rm acc}\frac{f Q_{\rm visc}}{\dot{m}_{\rm w}}}
		\propto \left(\frac{\alpha}{r}\right)^{1/2}.
	\end{equation}
	This scaling argument explains the robustness of the outflow velocity
	against variations in luminosity and the near linear correlation between the
	total kinetic power of the wind and the bolometric luminosity \citep[e.g.,
	T13, G15 and][]{2017A&A...601A.143F}, i.e.
	\begin{equation}
		\dot{E}_{\rm k}\propto \frac{1}{2}\int \dot{m}_{\rm w}v_{\rm out}^{2}{\rm d}
		r \propto \dot{m}_{\rm in}
	\end{equation}

	{We note that the constant velocity predicted here appears to deviate from the positive correlation between $v_{\rm out}$ and $\dot{m}_{\rm in}$ reported in some observations \citep[e.g.,][]{2017A&A...601A.143F, 2024A&A...687A.235G}. We will discuss the discrepancy in detail in Sec. \ref{sec:disscussion}.}

\begin{figure}
		\includegraphics[width=\columnwidth]{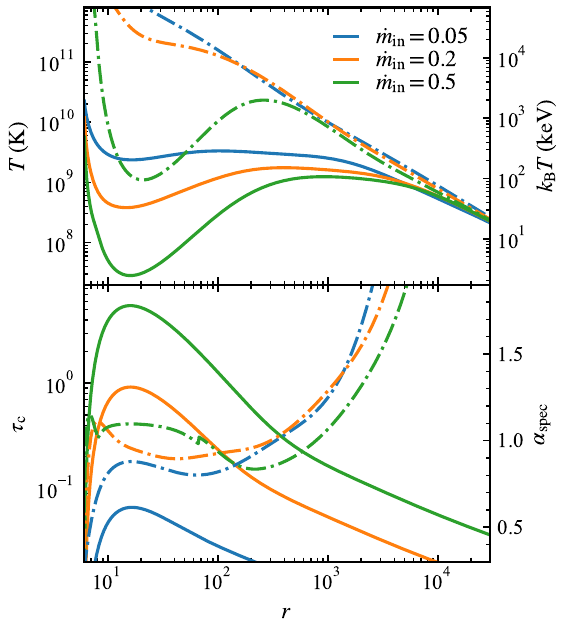}
		\caption{Radial structure of the corona for different accretion rate at
		the inner boundary $\dot{m}_{\rm in}$. The meanings of different line styles
		are the same as those in the Fig. \ref{fig:corona_profile}.}
		\label{fig:corona_profile_mdot}
	\end{figure}

	In Fig. \ref{fig:corona_profile_mdot}, we observe that as $\dot{m}_{\rm in}$
	increases, the coronal optical depth $\tau_{\rm c}$ rises, while both ion and
	electron temperatures $T_{\rm i}$ and $T_{\rm e}$ decrease. In the main X-ray
	emitting region ($10 < r < 100$), the electron temperature is highly sensitive
	to the accretion rate; for $\dot{m}_{\rm in}= 0.5$, $T_{\rm e}$ drops below $1
	0^{8}$ K due to efficient cooling. However, in the outer region ($r > 100$),
	which likely dominates the high-energy cutoff, $T_{\rm e}$ decreases more gradually.
	This trend is consistent with the observed anti-correlation between the high-energy
	cutoff temperature and the Eddington ratio in AGNs \citep[e.g.][]{2018MNRAS.480.1819R}.
	Furthermore, the spectral index in the main emitting region increases slowly
	with $\dot{m}_{\rm in}$, reproducing the well-known "brighter-softer"
	correlation \citep[e.g.][]{2006ApJ...646L..29S, 2008ApJ...682...81S}.

\begin{figure}
		\includegraphics[width=\columnwidth]{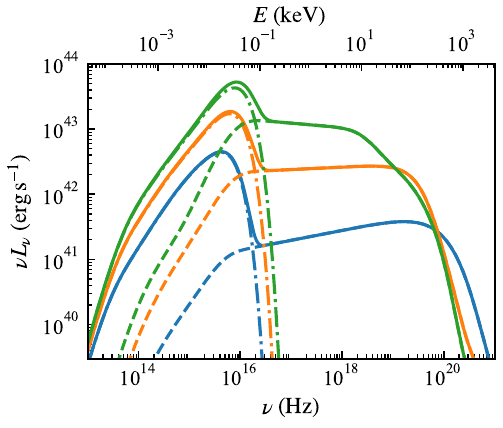}
		\caption{The spectra of the disc-corona systems for different accretion
		at the inner boundary $\dot{m}_{\rm in}$. The dash-dotted lines represent
		the blackbody component emitted by the accretion disc, the dashed lines
		represent the Compton component emitted by the corona, and the solid lines
		represent the total emission of the system. According to the order of
		luminosity from low to high, the curves of different colors correspond to
		$\dot{m}_{\rm in}= 0.05, 0.2$ and 0.5 respectively.}
		\label{fig:spectrum_mdot}
	\end{figure}

	Fig. \ref{fig:spectrum_mdot} displays the resulting broad-band spectra for different
	accretion rates. The total spectrum is calculated by integrating the local
	emission, where the Comptonized component is modeled using the analytical approximation
	of \citet{1994ApJ...434..570T} and normalized by the local energy
	dissipation (\ref{eq:compton_norm}). Although the local coronal properties ($T
	_{\rm e}$
	and $\tau_{\rm c}$) vary significantly with radius, the integrated X-ray
	spectra exhibit a clear power-law shape. The spectra naturally evolve to become
	softer and show a lower cutoff energy as the accretion rate increases, in
	excellent agreement with observational characteristics.

	\section{Discussion}
	\label{sec:disscussion}

	A challenge in accretion disc theory has been to explain the coexistence of
	powerful coronae and energetic outflows in the inner regions of black hole
	accretion flows, where radiation pressure dominates. The magnetic buoyancy
	model proposed by MF02 predicts a severe suppression of vertical energy
	transport in these regions. In their framework, the coronal fraction $f$
	drops to negligible levels ($f \ll 0.1$) in the radiation-pressure-dominant
	region, implying that the inner disc should be devoid of a strong corona or wind.
	C09 generalized this model by incorporating standard $\alpha$-viscosity
	scaling and the resultant energy transport is improved in the inner region. Our
	analysis demonstrates that in the inner region with high radiation pressure,
	the magnetic pressure may exceed the gas pressure and PBI will be excited
	\citep[e.g.,][]{1998MNRAS.297..929G}, which significantly enhances the vertical
	energy transport.
	{However, whether the PBI can survive in BH accretion disc environments still requires further research, as the strong turbulence in the disc may destroy the structure of photon bubbles before they can rise beyond the disc along with the flux tubes \citep[e.g.,][]{2005ApJ...624..267T}.}

	By incorporating the PBI-enhanced rising velocity (\ref{eq:velocity_pbi}) into
	the coronal energy budget, we find that the vertical energy flux is boosted
	by a factor of $\sqrt{\beta}$ compared to the C09 model and by orders of
	magnitude compared to the original MF02 predictions in the inner region. Our
	results show that with the PBI-driven enhancement, the inner disc has the
	energy to simultaneously sustain a hard X-ray corona and accelerate the UFOs
	to a high speed, validating the outflow-based corona scenario at the level of
	global energetics.

	Our model adopts the SI mechanism—specifically, a disc wind driven by
	vertical viscous torque in a weakly magnetized accretion disc—as the
	baseline for mass loading. This choice is applicable to the standard thin disc
	scenario, avoiding the requirement for the strong, large-scale magnetic fields
	typical of magnetocentrifugal models \citep[e.g.,][]{1982MNRAS.199..883B}.
	The fiducial parameters of $\dot{m}_{\rm in}\approx 0.2$ and
	$\alpha_{\rm w}\approx 0.01$ are found to yield a mass outflow rate profile
	that is in qualitative agreement with the observational UFO outflow rate
	derived by T13 and G15.

	Regarding the kinematics, we find that a relatively high mechanical efficiency,
	$f_{\rm acc}\gtrsim 0.5$, is required to accelerate the gas to the high velocities
	($v \sim 0.1c$) observed in UFOs within the inner region ($R \lesssim 100 R_{\rm
	g}$). This implies that the inner corona acts as a highly efficient gas
	accelerator, converting a major fraction of the buoyant magnetic energy into
	the bulk kinetic energy of the outflow. However, a discrepancy arises in the
	outer disc region ($R > 10^{3}R_{\rm g}$). As shown in Fig. \ref{fig:velocity_profile},
	even with high $f_{\rm acc}\rightarrow 1$, our magnetically driven model predicts
	velocities significantly lower than the observed value. This suggests a
	transition in the dominant driving mechanism. In the inner disc, the gas is
	highly ionized, suppressing line opacity \citep[e.g.,][]{2004ApJ...616..688P};
	thus, magnetic driving is likely the sole viable mechanism. As the outflow
	propagates to larger radii, the ionization parameter drops. Consequently, radiation
	pressure on spectral lines (line driving) becomes effective and likely takes
	over as the primary acceleration mechanism, boosting the wind velocity
	beyond the pure MHD prediction \citep[e.g.,][]{2000ApJ...543..686P}.

	It is also instructive to compare our scenario with other acceleration
	regimes. In super-Eddington sources, radiation pressure on the ionized
	electrons is likely the dominant driver of outflows, leading to a positive correlation
	between velocity and luminosity \citep[e.g.,][]{2003MNRAS.345..657K, 2017MNRAS.472L..15M, 2022ApJ...936..141C}.
	Conversely, in systems with extremely strong, large-scale magnetic fields, the
	magnetocentrifugal mechanism dominates \citep[e.g.,][]{1982MNRAS.199..883B}.
	However, for standard, weakly magnetized thin discs in the sub-Eddington
	regime, such extreme fields are difficult to generate \citep[e.g.,][]{1994MNRAS.267..235L,2013ApJ...767...30B}.
	In this context, the magnetic driving mechanism discussed in this work serves
	as a critical component, bridging the gap between the thermal driving at
	large radii and the gravitational potential well of the inner disc. Even if the
	magnetic field decays to equipartition with gas pressure at high altitudes,
	our model provides the essential kick to launch the high-ionization gas from
	the deep potential well, after which thermal, magnetic, and radiative forces
	may jointly sustain the coasting phase.

	{Observations of UFOs frequently reveal a positive correlation between the outflow velocity and the bolometric or X-ray luminosity \citep[e.g., G15;][]{2017A&A...601A.143F, 2024A&A...687A.235G}. This apparent discrepancy with our local coronal model can be naturally resolved by invoking a two-stage acceleration mechanism. If an outflow is primarily accelerated by the radiation pressure, its terminal velocity should be proportional to the square root of the luminosity, $v \propto L^{0.5}$ \citep[e.g.,][]{2017MNRAS.472L..15M}. However, observations often reveal a much shallower dependence, such as $v \propto L^{0.12}$ \citep[e.g.,][]{2024A&A...687A.235G}, though we note that empirical slopes can vary between studies due to observational selection effects. In the two-stage acceleration scenario, MHD-driven corona provides a massive kinematic foundation by accelerating the gas to a high initial base velocity. If the secondary acceleration phase is dominated by radiation pressure, the final terminal velocity would naturally exhibit a luminosity dependence that is significantly shallower than the $v \propto L^{0.5}$ relation predicted for purely radiation-driven winds.}

	In the UFO observations, the estimates of the UFO radius are often degenerate,
	relying on photoionization radius $R_{\rm max}= L / \xi N_{\rm H}$ or
	dynamical radius $R_{\rm min}= 2GM_{\rm BH}/ v^{2}$. T13 adopted a geometric
	mean $\sqrt{R_{\min}R_{\max}}$ to estimate the location, leading to a broad
	range of inferred radii. A recent work by \citet{2025arXiv251206077L}
	suggests that the dynamical radius of wind launching may extend to the ISCO,
	which indicates that the outflow-based corona may not be truncated but cover
	the entire inner disc in some sources. These observational findings align
	with our model, confirming that the innermost region, previously thought to
	be magnetically suppressed in MF02 model, can indeed host a powerful,
	outflowing corona essential for the sandwich geometry.

	In this paper, we also calculate the structure profiles of the corona with the
	radius and its thermodynamic evolution with the accretion rate. By solving
	the energy equations for ions and electrons, we derived radial profiles for the
	electron temperature and optical depth (see Fig. \ref{fig:corona_profile_mdot})
	that evolve naturally with the accretion rate. Specifically, our model
	reproduces the anti-correlation between the high-energy cutoff temperature
	$T_{\rm e, cut}$ and the Eddington ratio \citep[e.g.,][]{2018MNRAS.480.1819R},
	as well as the correlation between the spectral index and the Eddington ratio
	(i.e., the brighter-softer relation) in the statistical studies of AGNs \citep[e.g.,][]{2006ApJ...646L..29S,2008ApJ...682...81S, 2013MNRAS.433.2485B}.

	However, a closer inspection of the global energetics reveals a subtle
	tension when assuming fixed buoyancy efficiency $b_{0}$. In our current calculations,
	where the magnetic buoyancy efficiency factor $b_{0}$ is taken to be
	constant, the ratio of the X-ray luminosity in narrow band $2-10~{\rm keV}$ to
	the total bolometric luminosity $L_{\rm X}/L_{\rm bol}$ tends to increase as
	the accretion rate rises. (Due to the inverse correlation between ion/electron
	temperature and accretion rate, the fraction of the energy advection will
	decrease as the accretion rate increases.) This behavior contradicts the
	anti-correlation between $L_{\rm X}/ L_{\rm bol}$ observed in AGNs \citep[e.g.,][]{2007MNRAS.381.1235V, 2012MNRAS.425..623L, 2020A&A...636A..73D}.
	This discrepancy suggests that the efficiency of magnetic flux emergence is likely
	not constant but anti-correlated with the accretion rate.

	A physically compelling explanation lies in the geometry of the magnetic flux
	tubes. The characteristic wavelength of the Parker instability, which sets the
	length of the rising flux tubes, is proportional to the disc scale height, i.e.
	$8 .9H_{\rm d}\lesssim\lambda_{\rm P}\lesssim (3.5\beta + 6)H_{\rm d}$. As the
	accretion rate increases, the radiation-pressure-dominated inner disc
	becomes geometrically thicker. Consequently, the characteristic size of
	individual magnetic loops increases. For a given disc surface area, this
	leads to a reduction in the number density of independent flux tubes
	emerging from the magnetically active region ($N \propto 2\pi R / \lambda_{\rm
	P}$), which naturally decrease the effective covering factor $C_{\rm cover}$
	as the disc thickens. Therefore, we argue that the anti-correlation between $b
	_{0}$ and $\dot{m}$ may induce the anti-correlation between $L_{\rm cor}/ L_{\rm
	bol}$ (also $L_{\rm X}/ L_{\rm bol}$) and $\dot{m}$.

	In Sec. \ref{sec:varyingmdot}, we focused on moderate accretion rates $0.05\lesssim
	\dot{m}_{\rm in}\lesssim 0.5$ to avoid the possibility of state transitions.
	However, our outflow-based coronal model has significant implications for the
	global evolution of accretion flows.

	At low accretion rates ($\dot{m}\lesssim 0.01$), systems are known to
	transition from a standard thin disc to a radiatively inefficient, advection-dominated
	accretion flow (ADAF) \citep[e.g.,][]{1994ApJ...428L..13N, 1995ApJ...452..710N, 1997ApJ...489..865E}.
	In the context of the truncated disc model, the inner ADAF acts as the hot Comptonizing
	medium \citep[e.g.,][]{1997ApJ...489..865E,2007A&ARv..15....1D}. Although
	our model assumes a sandwich geometry, it is compatible with the truncation scenario.
	It is plausible that at low external accretion rates, the wind-driven mass loss
	causes the effective accretion rate in the inner region to drop below the critical
	value for the ADAF transition. In this picture, the low-luminosity corona might
	exhibit a two-phase structure: an outer outflow-based sandwich corona and an
	inner truncated ADAF.

	{Regarding the geometric size of the corona, our model predicts that both the vertical thickness and radial extent of the corona will expand with the accretion rate as the radiation-pressure-dominated region thickens and widens. This prediction presents an intriguing comparison with traditional paradigm. In the traditional truncated disc paradigm for XRBs, the inner ADAF shrinks as the accretion rate increases during the state transition \citep[e.g.,][]{remillardXRayPropertiesBlackHole2006}. This shrinking trend is supported by some X-ray reverberation mapping \citep[e.g.,][]{2015ApJ...814...50D, 2017MNRAS.471.1475D} and specific quasi-periodic oscillation (QPO) studies \citep[e.g.,][]{2009MNRAS.397L.101I, 2026JHEAp..5000513G}. However, this picture is currently debated. Reflection spectroscopy frequently reveals that the inner thin disc can extend to the ISCO even in hard states \citep[e.g.,][]{2008MNRAS.387.1489R, 2019Natur.565..198K}. Furthermore, recent observations have revealed that the coronal size positively correlates with luminosity in several sources, such as the AGN IRAS 13224-3809 \citep[e.g.,][]{2020NatAs...4..597A} and the microquasar MAXI J1820+070 \citep[e.g.,][]{2019Natur.565..198K}. Systematic spectral-timing studies also suggest a complex, non-monotonic evolution: as the accretion rate increases, the corona may initially shrink, but subsequently expand and stabilize at higher luminosities \citep[e.g.,][]{2022NatAs...6..577M, 2022MNRAS.512.2686Z, 2026MNRAS.546ag261Z}. Our PBI-enhanced slab corona model, which operates most effectively in the radiation-pressure-dominated regime at higher accretion rates, may account for the expanding phase of the corona. However, the exact evolution of the coronal geometry with the accretion rate still requires further investigation.}

	Conversely, when the accretion rate in XRB and AGN is very high, a strong corona
	may exist \citep[e.g.,][]{remillardXRayPropertiesBlackHole2006, 2006MNRAS.371.1216D}.
	In some ultraluminous sources, powerful outflows are also observed \citep[e.g.,][]{2016Natur.533...64P}.
	In such super-Eddington discs, the intense radiation pressure creates ideal conditions
	for vigorous PBI, maximizing the vertical transport of magnetic energy. Furthermore,
	the radiation pressure can also effectively drive the outflow acceleration. This
	alleviates the mechanical burden on the magnetic field; a much smaller
	fraction of magnetic energy $f_{\rm acc}$ is required for acceleration. Consequently,
	the bulk of the buoyant magnetic energy can be dissipated as heat, potentially
	explaining the coexistence of fast winds and luminous, strong coronal
	emission in these extreme sources.

	In this paper, the outflow-based corona model depends on the choice of the viscosity
	parameter $\alpha$. As indicated by the scaling relations in Section
	\ref{sec:varyingmdot}, the mass loading rate scales as $\dot{m}_{\rm w}\propto
	\alpha^{-1}$, while the terminal velocity scales as
	$v_{\rm out}\propto \alpha^{1/2}$. In this work, we adopted a fiducial value
	of $\alpha=0.1$, typical for the standard thin disc. Varying $\alpha$ would
	quantitatively shift the wind parameters needed but is unlikely to alter the
	qualitative trends and physical conclusions presented here.

	In this paper, to close the system of equations for the thermodynamic calculation,
	we adopted a prescribed kinematic profile rather than solving the full
	vertical momentum equation. We acknowledge that this is a simplification.
	The exact velocity profile depends on the specific process; for instance,
	self-similar magnetocentrifugal wind models typically predict a shallower
	scaling of $\rho\propto z^{-1.5}$ \citep[e.g.,][]{1982MNRAS.199..883B}. We have
	explored such density profile in our calculations but found that it results in
	excessive electron temperatures ($k_{\rm B}T_{\rm e}> 500$ keV) at low accretion
	rates that exceed observational constraints. Such temperatures would likely
	be physically regulated by other process, such as pair production,
	suggesting that additional cooling mechanisms omitted in our current energy
	equation would become significant in that regime. In contrast, the adopted
	quadratic profile captures the essential accelerating nature of the flow
	while naturally reproducing electron temperatures consistent with hard X-ray
	observations, serving as a fiducial approximation for this study. In a
	forthcoming study (Paper II), we will relax this assumption by solving the coupled
	inter-relations between the magnetic pressure, gas density, and velocity field
	, which will provide a self-consistent vertical structure of the corona.

	\section*{Acknowledgements}

	I am deeply grateful to the reviewer for his/her valuable and insightful
	comments, which helped to greatly improve and refine this work. I am also
	grateful to Prof. Xinwu Cao for his continuous support and helpful discussions
	throughout this work and Prof. Agnieszka Janiuk for her valuable comments
	and useful discussions. This work is supported by the NSFC (12533005, 12233007,
	12347103, and 12361131579), the science research grants from the China Manned
	Space Project with CMS-CSST-2025-A07, and the fundamental research fund for
	Chinese central universities (Zhejiang University).

	\section*{Data Availability}

	The data and code underlying this article will be shared on reasonable request
	to the corresponding author.


	\bibliographystyle{mnras}
	\bibliography{example} 

\begin{thebibliography}{}
\makeatletter
\relax
\def\mn@urlcharsother{\let\do\@makeother \do\$\do\&\do\#\do\^\do\_\do\%\do\~}
\def\mn@doi{\begingroup\mn@urlcharsother \@ifnextchar [ {\mn@doi@}
  {\mn@doi@[]}}
\def\mn@doi@[#1]#2{\def\@tempa{#1}\ifx\@tempa\@empty \href
  {http://dx.doi.org/#2} {doi:#2}\else \href {http://dx.doi.org/#2} {#1}\fi
  \endgroup}
\def\mn@eprint#1#2{\mn@eprint@#1:#2::\@nil}
\def\mn@eprint@arXiv#1{\href {http://arxiv.org/abs/#1} {{\tt arXiv:#1}}}
\def\mn@eprint@dblp#1{\href {http://dblp.uni-trier.de/rec/bibtex/#1.xml}
  {dblp:#1}}
\def\mn@eprint@#1:#2:#3:#4\@nil{\def\@tempa {#1}\def\@tempb {#2}\def\@tempc
  {#3}\ifx \@tempc \@empty \let \@tempc \@tempb \let \@tempb \@tempa \fi \ifx
  \@tempb \@empty \def\@tempb {arXiv}\fi \@ifundefined
  {mn@eprint@\@tempb}{\@tempb:\@tempc}{\expandafter \expandafter \csname
  mn@eprint@\@tempb\endcsname \expandafter{\@tempc}}}

\bibitem[\protect\citeauthoryear{{Alston} et~al.,}{{Alston}
  et~al.}{2020}]{2020NatAs...4..597A}
{Alston} W.~N.,  et~al., 2020, \mn@doi [Nature Astronomy]
  {10.1038/s41550-019-1002-x}, \href
  {https://ui.adsabs.harvard.edu/abs/2020NatAs...4..597A} {4, 597}

\bibitem[\protect\citeauthoryear{{Arons}}{{Arons}}{1992}]{1992ApJ...388..561A}
{Arons} J.,  1992, \mn@doi [\apj] {10.1086/171174}, \href
  {https://ui.adsabs.harvard.edu/abs/1992ApJ...388..561A} {388, 561}

\bibitem[\protect\citeauthoryear{{Bai} \& {Stone}}{{Bai} \&
  {Stone}}{2013}]{2013ApJ...767...30B}
{Bai} X.-N.,  {Stone} J.~M.,  2013, \mn@doi [\apj]
  {10.1088/0004-637X/767/1/30}, \href
  {https://ui.adsabs.harvard.edu/abs/2013ApJ...767...30B} {767, 30}

\bibitem[\protect\citeauthoryear{{Balbus} \& {Hawley}}{{Balbus} \&
  {Hawley}}{1991}]{BH91}
{Balbus} S.~A.,  {Hawley} J.~F.,  1991, \mn@doi [\apj] {10.1086/170270}, \href
  {https://ui.adsabs.harvard.edu/abs/1991ApJ...376..214B} {376, 214}

\bibitem[\protect\citeauthoryear{{Balbus} \& {Hawley}}{{Balbus} \&
  {Hawley}}{1998}]{1998RvMP...70....1B}
{Balbus} S.~A.,  {Hawley} J.~F.,  1998, \mn@doi [Reviews of Modern Physics]
  {10.1103/RevModPhys.70.1}, \href
  {https://ui.adsabs.harvard.edu/abs/1998RvMP...70....1B} {70, 1}

\bibitem[\protect\citeauthoryear{{Begelman}}{{Begelman}}{2001}]{2001ApJ...551..897B}
{Begelman} M.~C.,  2001, \mn@doi [\apj] {10.1086/320240}, \href
  {https://ui.adsabs.harvard.edu/abs/2001ApJ...551..897B} {551, 897}

\bibitem[\protect\citeauthoryear{{Blackman}, {Penna}  \&
  {Varni{\`e}re}}{{Blackman} et~al.}{2008}]{2008NewA...13..244B}
{Blackman} E.~G.,  {Penna} R.~F.,   {Varni{\`e}re} P.,  2008, \mn@doi [\na]
  {10.1016/j.newast.2007.10.004}, \href
  {https://ui.adsabs.harvard.edu/abs/2008NewA...13..244B} {13, 244}

\bibitem[\protect\citeauthoryear{{Blaes} \& {Socrates}}{{Blaes} \&
  {Socrates}}{2001}]{2001ApJ...553..987B}
{Blaes} O.,  {Socrates} A.,  2001, \mn@doi [\apj] {10.1086/320968}, \href
  {https://ui.adsabs.harvard.edu/abs/2001ApJ...553..987B} {553, 987}

\bibitem[\protect\citeauthoryear{{Blaes} \& {Socrates}}{{Blaes} \&
  {Socrates}}{2003}]{2003ApJ...596..509B}
{Blaes} O.,  {Socrates} A.,  2003, \mn@doi [\apj] {10.1086/377637}, \href
  {https://ui.adsabs.harvard.edu/abs/2003ApJ...596..509B} {596, 509}

\bibitem[\protect\citeauthoryear{{Blaes}, {Hirose}  \& {Krolik}}{{Blaes}
  et~al.}{2007}]{2007ApJ...664.1057B}
{Blaes} O.,  {Hirose} S.,   {Krolik} J.~H.,  2007, \mn@doi [\apj]
  {10.1086/519516}, \href
  {https://ui.adsabs.harvard.edu/abs/2007ApJ...664.1057B} {664, 1057}

\bibitem[\protect\citeauthoryear{{Blandford} \& {Payne}}{{Blandford} \&
  {Payne}}{1982}]{1982MNRAS.199..883B}
{Blandford} R.~D.,  {Payne} D.~G.,  1982, \mn@doi [\mnras]
  {10.1093/mnras/199.4.883}, \href
  {https://ui.adsabs.harvard.edu/abs/1982MNRAS.199..883B} {199, 883}

\bibitem[\protect\citeauthoryear{{Brightman} et~al.,}{{Brightman}
  et~al.}{2013}]{2013MNRAS.433.2485B}
{Brightman} M.,  et~al., 2013, \mn@doi [\mnras] {10.1093/mnras/stt920}, \href
  {https://ui.adsabs.harvard.edu/abs/2013MNRAS.433.2485B} {433, 2485}

\bibitem[\protect\citeauthoryear{{Cao}}{{Cao}}{2009}]{C09}
{Cao} X.,  2009, \mn@doi [\mnras] {10.1111/j.1365-2966.2008.14347.x}, \href
  {https://ui.adsabs.harvard.edu/abs/2009MNRAS.394..207C} {394, 207}

\bibitem[\protect\citeauthoryear{{Cao} \& {Gu}}{{Cao} \&
  {Gu}}{2022}]{2022ApJ...936..141C}
{Cao} X.,  {Gu} W.-M.,  2022, \mn@doi [\apj] {10.3847/1538-4357/ac8980}, \href
  {https://ui.adsabs.harvard.edu/abs/2022ApJ...936..141C} {936, 141}

\bibitem[\protect\citeauthoryear{{Chartas}, {Kochanek}, {Dai}, {Poindexter}  \&
  {Garmire}}{{Chartas} et~al.}{2009}]{2009ApJ...693..174C}
{Chartas} G.,  {Kochanek} C.~S.,  {Dai} X.,  {Poindexter} S.,   {Garmire} G.,
  2009, \mn@doi [\apj] {10.1088/0004-637X/693/1/174}, \href
  {https://ui.adsabs.harvard.edu/abs/2009ApJ...693..174C} {693, 174}

\bibitem[\protect\citeauthoryear{{Chartas} et~al.,}{{Chartas}
  et~al.}{2016}]{2016AN....337..356C}
{Chartas} G.,  et~al., 2016, \mn@doi [Astronomische Nachrichten]
  {10.1002/asna.201612313}, \href
  {https://ui.adsabs.harvard.edu/abs/2016AN....337..356C} {337, 356}

\bibitem[\protect\citeauthoryear{{De Marco}, {Ponti}, {Cappi}, {Dadina},
  {Uttley}, {Cackett}, {Fabian}  \& {Miniutti}}{{De Marco}
  et~al.}{2013}]{2013MNRAS.431.2441D}
{De Marco} B.,  {Ponti} G.,  {Cappi} M.,  {Dadina} M.,  {Uttley} P.,  {Cackett}
  E.~M.,  {Fabian} A.~C.,   {Miniutti} G.,  2013, \mn@doi [\mnras]
  {10.1093/mnras/stt339}, \href
  {https://ui.adsabs.harvard.edu/abs/2013MNRAS.431.2441D} {431, 2441}

\bibitem[\protect\citeauthoryear{{De Marco}, {Ponti}, {Mu{\~n}oz-Darias}  \&
  {Nandra}}{{De Marco} et~al.}{2015}]{2015ApJ...814...50D}
{De Marco} B.,  {Ponti} G.,  {Mu{\~n}oz-Darias} T.,   {Nandra} K.,  2015,
  \mn@doi [\apj] {10.1088/0004-637X/814/1/50}, \href
  {https://ui.adsabs.harvard.edu/abs/2015ApJ...814...50D} {814, 50}

\bibitem[\protect\citeauthoryear{{De Marco} et~al.,}{{De Marco}
  et~al.}{2017}]{2017MNRAS.471.1475D}
{De Marco} B.,  et~al., 2017, \mn@doi [\mnras] {10.1093/mnras/stx1649}, \href
  {https://ui.adsabs.harvard.edu/abs/2017MNRAS.471.1475D} {471, 1475}

\bibitem[\protect\citeauthoryear{{Di Matteo}}{{Di
  Matteo}}{1998}]{1998MNRAS.299L..15D}
{Di Matteo} T.,  1998, \mn@doi [\mnras] {10.1046/j.1365-8711.1998.01950.x},
  \href {https://ui.adsabs.harvard.edu/abs/1998MNRAS.299L..15D} {299, L15}

\bibitem[\protect\citeauthoryear{{Done} \& {Kubota}}{{Done} \&
  {Kubota}}{2006}]{2006MNRAS.371.1216D}
{Done} C.,  {Kubota} A.,  2006, \mn@doi [\mnras]
  {10.1111/j.1365-2966.2006.10737.x}, \href
  {https://ui.adsabs.harvard.edu/abs/2006MNRAS.371.1216D} {371, 1216}

\bibitem[\protect\citeauthoryear{{Done}, {Gierli{\'n}ski}  \& {Kubota}}{{Done}
  et~al.}{2007}]{2007A&ARv..15....1D}
{Done} C.,  {Gierli{\'n}ski} M.,   {Kubota} A.,  2007, \mn@doi [\aapr]
  {10.1007/s00159-007-0006-1}, \href
  {https://ui.adsabs.harvard.edu/abs/2007A&ARv..15....1D} {15, 1}

\bibitem[\protect\citeauthoryear{{Duras} et~al.,}{{Duras}
  et~al.}{2020}]{2020A&A...636A..73D}
{Duras} F.,  et~al., 2020, \mn@doi [\aap] {10.1051/0004-6361/201936817}, \href
  {https://ui.adsabs.harvard.edu/abs/2020A&A...636A..73D} {636, A73}

\bibitem[\protect\citeauthoryear{{Eastwood}, {Phan}, {Drake}, {Shay}, {Borg},
  {Lavraud}  \& {Taylor}}{{Eastwood} et~al.}{2013}]{2013PhRvL.110v5001E}
{Eastwood} J.~P.,  {Phan} T.~D.,  {Drake} J.~F.,  {Shay} M.~A.,  {Borg} A.~L.,
  {Lavraud} B.,   {Taylor} M.~G.~G.~T.,  2013, \mn@doi [\prl]
  {10.1103/PhysRevLett.110.225001}, \href
  {https://ui.adsabs.harvard.edu/abs/2013PhRvL.110v5001E} {110, 225001}

\bibitem[\protect\citeauthoryear{{Esin}, {McClintock}  \& {Narayan}}{{Esin}
  et~al.}{1997}]{1997ApJ...489..865E}
{Esin} A.~A.,  {McClintock} J.~E.,   {Narayan} R.,  1997, \mn@doi [\apj]
  {10.1086/304829}, \href
  {https://ui.adsabs.harvard.edu/abs/1997ApJ...489..865E} {489, 865}

\bibitem[\protect\citeauthoryear{{Fabian}, {Lohfink}, {Kara}, {Parker},
  {Vasudevan}  \& {Reynolds}}{{Fabian} et~al.}{2015}]{2015MNRAS.451.4375F}
{Fabian} A.~C.,  {Lohfink} A.,  {Kara} E.,  {Parker} M.~L.,  {Vasudevan} R.,
  {Reynolds} C.~S.,  2015, \mn@doi [\mnras] {10.1093/mnras/stv1218}, \href
  {https://ui.adsabs.harvard.edu/abs/2015MNRAS.451.4375F} {451, 4375}

\bibitem[\protect\citeauthoryear{{Fiore} et~al.,}{{Fiore}
  et~al.}{2017}]{2017A&A...601A.143F}
{Fiore} F.,  et~al., 2017, \mn@doi [\aap] {10.1051/0004-6361/201629478}, \href
  {https://ui.adsabs.harvard.edu/abs/2017A&A...601A.143F} {601, A143}

\bibitem[\protect\citeauthoryear{{Frank}, {King}  \& {Raine}}{{Frank}
  et~al.}{2002}]{2002apa..book.....F}
{Frank} J.,  {King} A.,   {Raine} D.~J.,  2002, {Accretion Power in
  Astrophysics: Third Edition}

\bibitem[\protect\citeauthoryear{{Galeev}, {Rosner}  \& {Vaiana}}{{Galeev}
  et~al.}{1979}]{GS79}
{Galeev} A.~A.,  {Rosner} R.,   {Vaiana} G.~S.,  1979, \mn@doi [\apj]
  {10.1086/156957}, \href
  {https://ui.adsabs.harvard.edu/abs/1979ApJ...229..318G} {229, 318}

\bibitem[\protect\citeauthoryear{{Gammie}}{{Gammie}}{1998}]{1998MNRAS.297..929G}
{Gammie} C.~F.,  1998, \mn@doi [\mnras] {10.1046/j.1365-8711.1998.01571.x},
  \href {https://ui.adsabs.harvard.edu/abs/1998MNRAS.297..929G} {297, 929}

\bibitem[\protect\citeauthoryear{{Garc{\'\i}a}, {Dauser}, {Reynolds},
  {Kallman}, {McClintock}, {Wilms}  \& {Eikmann}}{{Garc{\'\i}a}
  et~al.}{2013}]{2013ApJ...768..146G}
{Garc{\'\i}a} J.,  {Dauser} T.,  {Reynolds} C.~S.,  {Kallman} T.~R.,
  {McClintock} J.~E.,  {Wilms} J.,   {Eikmann} W.,  2013, \mn@doi [\apj]
  {10.1088/0004-637X/768/2/146}, \href
  {https://ui.adsabs.harvard.edu/abs/2013ApJ...768..146G} {768, 146}

\bibitem[\protect\citeauthoryear{{Ghisellini}}{{Ghisellini}}{2013}]{2013LNP...873.....G}
{Ghisellini} G.,  2013, {Radiative Processes in High Energy Astrophysics}.
 Vol. 873, \mn@doi{10.1007/978-3-319-00612-3, }

\bibitem[\protect\citeauthoryear{{Ghising}, {Subba}, {Tobrej}, {Rai}  \&
  {Paul}}{{Ghising} et~al.}{2026}]{2026JHEAp..5000513G}
{Ghising} M.,  {Subba} N.,  {Tobrej} M.,  {Rai} B.,   {Paul} B.~C.,  2026,
  \mn@doi [Journal of High Energy Astrophysics] {10.1016/j.jheap.2025.100513},
  \href {https://ui.adsabs.harvard.edu/abs/2026JHEAp..5000513G} {50, 100513}

\bibitem[\protect\citeauthoryear{{Giampaoli}, {Rodrigues}, {Rodrigues}  \&
  {Mendon{\c{c}}a}}{{Giampaoli} et~al.}{2021}]{2021NatCo..12.3240G}
{Giampaoli} R.,  {Rodrigues} J.~D.,  {Rodrigues} J.-A.,   {Mendon{\c{c}}a}
  J.~T.,  2021, \mn@doi [Nature Communications] {10.1038/s41467-021-23493-2},
  \href {https://ui.adsabs.harvard.edu/abs/2021NatCo..12.3240G} {12, 3240}

\bibitem[\protect\citeauthoryear{{Gianolli} et~al.,}{{Gianolli}
  et~al.}{2023}]{2023MNRAS.523.4468G}
{Gianolli} V.~E.,  et~al., 2023, \mn@doi [\mnras] {10.1093/mnras/stad1697},
  \href {https://ui.adsabs.harvard.edu/abs/2023MNRAS.523.4468G} {523, 4468}

\bibitem[\protect\citeauthoryear{{Gianolli} et~al.,}{{Gianolli}
  et~al.}{2024}]{2024A&A...687A.235G}
{Gianolli} V.~E.,  et~al., 2024, \mn@doi [\aap] {10.1051/0004-6361/202348908},
  \href {https://ui.adsabs.harvard.edu/abs/2024A&A...687A.235G} {687, A235}

\bibitem[\protect\citeauthoryear{{Giz} \& {Shu}}{{Giz} \&
  {Shu}}{1993}]{1993ApJ...404..185G}
{Giz} A.~T.,  {Shu} F.~H.,  1993, \mn@doi [\apj] {10.1086/172267}, \href
  {https://ui.adsabs.harvard.edu/abs/1993ApJ...404..185G} {404, 185}

\bibitem[\protect\citeauthoryear{{Gofford}, {Reeves}, {Tombesi}, {Braito},
  {Turner}, {Miller}  \& {Cappi}}{{Gofford} et~al.}{2013}]{2013MNRAS.430...60G}
{Gofford} J.,  {Reeves} J.~N.,  {Tombesi} F.,  {Braito} V.,  {Turner} T.~J.,
  {Miller} L.,   {Cappi} M.,  2013, \mn@doi [\mnras] {10.1093/mnras/sts481},
  \href {https://ui.adsabs.harvard.edu/abs/2013MNRAS.430...60G} {430, 60}

\bibitem[\protect\citeauthoryear{{Gofford}, {Reeves}, {McLaughlin}, {Braito},
  {Turner}, {Tombesi}  \& {Cappi}}{{Gofford}
  et~al.}{2015}]{2015MNRAS.451.4169G}
{Gofford} J.,  {Reeves} J.~N.,  {McLaughlin} D.~E.,  {Braito} V.,  {Turner}
  T.~J.,  {Tombesi} F.,   {Cappi} M.,  2015, \mn@doi [\mnras]
  {10.1093/mnras/stv1207}, \href
  {https://ui.adsabs.harvard.edu/abs/2015MNRAS.451.4169G} {451, 4169}

\bibitem[\protect\citeauthoryear{Goldstein}{Goldstein}{1938}]{goldstein1938modern}
Goldstein S.,  1938, Modern developments in fluid dynamics: an account of
  theory and experiment relating to boundary layers, turbulent motion and
  wakes.
 Vol. 2, Clarendon Press

\bibitem[\protect\citeauthoryear{{Haardt} \& {Maraschi}}{{Haardt} \&
  {Maraschi}}{1991}]{HM91}
{Haardt} F.,  {Maraschi} L.,  1991, \mn@doi [\apjl] {10.1086/186171}, \href
  {https://ui.adsabs.harvard.edu/abs/1991ApJ...380L..51H} {380, L51}

\bibitem[\protect\citeauthoryear{{Haardt} \& {Maraschi}}{{Haardt} \&
  {Maraschi}}{1993}]{HM93}
{Haardt} F.,  {Maraschi} L.,  1993, \mn@doi [\apj] {10.1086/173020}, \href
  {https://ui.adsabs.harvard.edu/abs/1993ApJ...413..507H} {413, 507}

\bibitem[\protect\citeauthoryear{{Haardt}, {Maraschi}  \&
  {Ghisellini}}{{Haardt} et~al.}{1994}]{1994ApJ...432L..95H}
{Haardt} F.,  {Maraschi} L.,   {Ghisellini} G.,  1994, \mn@doi [\apjl]
  {10.1086/187520}, \href
  {https://ui.adsabs.harvard.edu/abs/1994ApJ...432L..95H} {432, L95}

\bibitem[\protect\citeauthoryear{{Ingram}, {Done}  \& {Fragile}}{{Ingram}
  et~al.}{2009}]{2009MNRAS.397L.101I}
{Ingram} A.,  {Done} C.,   {Fragile} P.~C.,  2009, \mn@doi [\mnras]
  {10.1111/j.1745-3933.2009.00693.x}, \href
  {https://ui.adsabs.harvard.edu/abs/2009MNRAS.397L.101I} {397, L101}

\bibitem[\protect\citeauthoryear{{Jacquemin-Ide}, {Ferreira}  \&
  {Lesur}}{{Jacquemin-Ide} et~al.}{2019}]{2019MNRAS.490.3112J}
{Jacquemin-Ide} J.,  {Ferreira} J.,   {Lesur} G.,  2019, \mn@doi [\mnras]
  {10.1093/mnras/stz2749}, \href
  {https://ui.adsabs.harvard.edu/abs/2019MNRAS.490.3112J} {490, 3112}

\bibitem[\protect\citeauthoryear{{Jacquemin-Ide}, {Lesur}  \&
  {Ferreira}}{{Jacquemin-Ide} et~al.}{2021}]{2021A&A...647A.192J}
{Jacquemin-Ide} J.,  {Lesur} G.,   {Ferreira} J.,  2021, \mn@doi [\aap]
  {10.1051/0004-6361/202039322}, \href
  {https://ui.adsabs.harvard.edu/abs/2021A&A...647A.192J} {647, A192}

\bibitem[\protect\citeauthoryear{{Jiang}, {Stone}  \& {Davis}}{{Jiang}
  et~al.}{2014}]{2014ApJ...796..106J}
{Jiang} Y.-F.,  {Stone} J.~M.,   {Davis} S.~W.,  2014, \mn@doi [\apj]
  {10.1088/0004-637X/796/2/106}, \href
  {https://ui.adsabs.harvard.edu/abs/2014ApJ...796..106J} {796, 106}

\bibitem[\protect\citeauthoryear{{Jiang}, {Stone}  \& {Davis}}{{Jiang}
  et~al.}{2019}]{2019ApJ...880...67J}
{Jiang} Y.-F.,  {Stone} J.~M.,   {Davis} S.~W.,  2019, \mn@doi [\apj]
  {10.3847/1538-4357/ab29ff}, \href
  {https://ui.adsabs.harvard.edu/abs/2019ApJ...880...67J} {880, 67}

\bibitem[\protect\citeauthoryear{{Kara} et~al.,}{{Kara}
  et~al.}{2019}]{2019Natur.565..198K}
{Kara} E.,  et~al., 2019, \mn@doi [\nat] {10.1038/s41586-018-0803-x}, \href
  {https://ui.adsabs.harvard.edu/abs/2019Natur.565..198K} {565, 198}

\bibitem[\protect\citeauthoryear{{Kato}, {Fukue}  \& {Mineshige}}{{Kato}
  et~al.}{2008}]{2008bhad.book.....K}
{Kato} S.,  {Fukue} J.,   {Mineshige} S.,  2008, {Black-Hole Accretion Disks
  --- Towards a New Paradigm ---}

\bibitem[\protect\citeauthoryear{{King} \& {Pounds}}{{King} \&
  {Pounds}}{2003}]{2003MNRAS.345..657K}
{King} A.~R.,  {Pounds} K.~A.,  2003, \mn@doi [\mnras]
  {10.1046/j.1365-8711.2003.06980.x}, \href
  {https://ui.adsabs.harvard.edu/abs/2003MNRAS.345..657K} {345, 657}

\bibitem[\protect\citeauthoryear{{King}, {Pringle}  \& {Livio}}{{King}
  et~al.}{2007}]{2007MNRAS.376.1740K}
{King} A.~R.,  {Pringle} J.~E.,   {Livio} M.,  2007, \mn@doi [\mnras]
  {10.1111/j.1365-2966.2007.11556.x}, \href
  {https://ui.adsabs.harvard.edu/abs/2007MNRAS.376.1740K} {376, 1740}

\bibitem[\protect\citeauthoryear{{Kippenhahn} \& {Weigert}}{{Kippenhahn} \&
  {Weigert}}{1990}]{1990sse..book.....K}
{Kippenhahn} R.,  {Weigert} A.,  1990, {Stellar Structure and Evolution}

\bibitem[\protect\citeauthoryear{{Krawczynski} et~al.,}{{Krawczynski}
  et~al.}{2022}]{2022Sci...378..650K}
{Krawczynski} H.,  et~al., 2022, \mn@doi [Science] {10.1126/science.add5399},
  \href {https://ui.adsabs.harvard.edu/abs/2022Sci...378..650K} {378, 650}

\bibitem[\protect\citeauthoryear{{Krolik} \& {Kriss}}{{Krolik} \&
  {Kriss}}{1995}]{1995ApJ...447..512K}
{Krolik} J.~H.,  {Kriss} G.~A.,  1995, \mn@doi [\apj] {10.1086/175896}, \href
  {https://ui.adsabs.harvard.edu/abs/1995ApJ...447..512K} {447, 512}

\bibitem[\protect\citeauthoryear{{Laurenti} et~al.,}{{Laurenti}
  et~al.}{2025}]{2025arXiv251206077L}
{Laurenti} M.,  et~al., 2025, arXiv e-prints, \href
  {https://ui.adsabs.harvard.edu/abs/2025arXiv251206077L} {p. arXiv:2512.06077}

\bibitem[\protect\citeauthoryear{{Liang} \& {Price}}{{Liang} \&
  {Price}}{1977}]{1977ApJ...218..247L}
{Liang} E.~P.~T.,  {Price} R.~H.,  1977, \mn@doi [\apj] {10.1086/155677}, \href
  {https://ui.adsabs.harvard.edu/abs/1977ApJ...218..247L} {218, 247}

\bibitem[\protect\citeauthoryear{{Lightman} \& {Eardley}}{{Lightman} \&
  {Eardley}}{1974}]{1974ApJ...187L...1L}
{Lightman} A.~P.,  {Eardley} D.~M.,  1974, \mn@doi [\apjl] {10.1086/181377},
  \href {https://ui.adsabs.harvard.edu/abs/1974ApJ...187L...1L} {187, L1}

\bibitem[\protect\citeauthoryear{{Lubow}, {Papaloizou}  \& {Pringle}}{{Lubow}
  et~al.}{1994}]{1994MNRAS.267..235L}
{Lubow} S.~H.,  {Papaloizou} J.~C.~B.,   {Pringle} J.~E.,  1994, \mn@doi
  [\mnras] {10.1093/mnras/267.2.235}, \href
  {https://ui.adsabs.harvard.edu/abs/1994MNRAS.267..235L} {267, 235}

\bibitem[\protect\citeauthoryear{{Lusso} et~al.,}{{Lusso}
  et~al.}{2012}]{2012MNRAS.425..623L}
{Lusso} E.,  et~al., 2012, \mn@doi [\mnras] {10.1111/j.1365-2966.2012.21513.x},
  \href {https://ui.adsabs.harvard.edu/abs/2012MNRAS.425..623L} {425, 623}

\bibitem[\protect\citeauthoryear{{Lynden-Bell}}{{Lynden-Bell}}{1969}]{lynden-bellGalacticNucleiCollapsed1969a}
{Lynden-Bell} D.,  1969, \mn@doi [\nat] {10.1038/223690a0}, 223, 690

\bibitem[\protect\citeauthoryear{{Lynden-Bell}}{{Lynden-Bell}}{1996}]{1996MNRAS.279..389L}
{Lynden-Bell} D.,  1996, \mn@doi [\mnras] {10.1093/mnras/279.2.389}, \href
  {https://ui.adsabs.harvard.edu/abs/1996MNRAS.279..389L} {279, 389}

\bibitem[\protect\citeauthoryear{{Lynden-Bell}}{{Lynden-Bell}}{2003}]{2003MNRAS.341.1360L}
{Lynden-Bell} D.,  2003, \mn@doi [\mnras] {10.1046/j.1365-8711.2003.06506.x},
  \href {https://ui.adsabs.harvard.edu/abs/2003MNRAS.341.1360L} {341, 1360}

\bibitem[\protect\citeauthoryear{{Markoff}, {Nowak}  \& {Wilms}}{{Markoff}
  et~al.}{2005}]{2005ApJ...635.1203M}
{Markoff} S.,  {Nowak} M.~A.,   {Wilms} J.,  2005, \mn@doi [\apj]
  {10.1086/497628}, \href
  {https://ui.adsabs.harvard.edu/abs/2005ApJ...635.1203M} {635, 1203}

\bibitem[\protect\citeauthoryear{{Matsumoto} \& {Shibata}}{{Matsumoto} \&
  {Shibata}}{1991}]{1991IAUS..144..429M}
{Matsumoto} R.,  {Shibata} K.,  1991, in {Bloemen} H.,  ed.,  IAU Symposium
  Vol. 144, The Interstellar Disk-Halo Connection in Galaxies. p.~429

\bibitem[\protect\citeauthoryear{{Matt}, {Perola}  \& {Piro}}{{Matt}
  et~al.}{1991}]{1991A&A...247...25M}
{Matt} G.,  {Perola} G.~C.,   {Piro} L.,  1991, \aap, \href
  {https://ui.adsabs.harvard.edu/abs/1991A&A...247...25M} {247, 25}

\bibitem[\protect\citeauthoryear{{Matzeu}, {Reeves}, {Braito}, {Nardini},
  {McLaughlin}, {Lobban}, {Tombesi}  \& {Costa}}{{Matzeu}
  et~al.}{2017}]{2017MNRAS.472L..15M}
{Matzeu} G.~A.,  {Reeves} J.~N.,  {Braito} V.,  {Nardini} E.,  {McLaughlin}
  D.~E.,  {Lobban} A.~P.,  {Tombesi} F.,   {Costa} M.~T.,  2017, \mn@doi
  [\mnras] {10.1093/mnrasl/slx129}, \href
  {https://ui.adsabs.harvard.edu/abs/2017MNRAS.472L..15M} {472, L15}

\bibitem[\protect\citeauthoryear{{M{\'e}ndez}, {Karpouzas}, {Garc{\'\i}a},
  {Zhang}, {Zhang}, {Belloni}  \& {Altamirano}}{{M{\'e}ndez}
  et~al.}{2022}]{2022NatAs...6..577M}
{M{\'e}ndez} M.,  {Karpouzas} K.,  {Garc{\'\i}a} F.,  {Zhang} L.,  {Zhang} Y.,
  {Belloni} T.~M.,   {Altamirano} D.,  2022, \mn@doi [Nature Astronomy]
  {10.1038/s41550-022-01617-y}, \href
  {https://ui.adsabs.harvard.edu/abs/2022NatAs...6..577M} {6, 577}

\bibitem[\protect\citeauthoryear{{Merloni} \& {Fabian}}{{Merloni} \&
  {Fabian}}{2002}]{MF02}
{Merloni} A.,  {Fabian} A.~C.,  2002, \mn@doi [\mnras]
  {10.1046/j.1365-8711.2002.05288.x}, \href
  {https://ui.adsabs.harvard.edu/abs/2002MNRAS.332..165M} {332, 165}

\bibitem[\protect\citeauthoryear{{Miller} \& {Stone}}{{Miller} \&
  {Stone}}{2000}]{MS00}
{Miller} K.~A.,  {Stone} J.~M.,  2000, \mn@doi [\apj] {10.1086/308736}, \href
  {https://ui.adsabs.harvard.edu/abs/2000ApJ...534..398M} {534, 398}

\bibitem[\protect\citeauthoryear{{Miller}, {Raymond}, {Fabian}, {Steeghs},
  {Homan}, {Reynolds}, {van der Klis}  \& {Wijnands}}{{Miller}
  et~al.}{2006}]{2006Natur.441..953M}
{Miller} J.~M.,  {Raymond} J.,  {Fabian} A.,  {Steeghs} D.,  {Homan} J.,
  {Reynolds} C.,  {van der Klis} M.,   {Wijnands} R.,  2006, \mn@doi [\nat]
  {10.1038/nature04912}, \href
  {https://ui.adsabs.harvard.edu/abs/2006Natur.441..953M} {441, 953}

\bibitem[\protect\citeauthoryear{{Mosquera}, {Kochanek}, {Chen}, {Dai},
  {Blackburne}  \& {Chartas}}{{Mosquera} et~al.}{2013}]{2013ApJ...769...53M}
{Mosquera} A.~M.,  {Kochanek} C.~S.,  {Chen} B.,  {Dai} X.,  {Blackburne}
  J.~A.,   {Chartas} G.,  2013, \mn@doi [\apj] {10.1088/0004-637X/769/1/53},
  \href {https://ui.adsabs.harvard.edu/abs/2013ApJ...769...53M} {769, 53}

\bibitem[\protect\citeauthoryear{{Mutie}, {del Palacio}, {Beswick},
  {Williams-Baldwin}, {Gallimore}, {Gallagher}, {Aalto}  \& {Baki}}{{Mutie}
  et~al.}{2025}]{2025MNRAS.539..808M}
{Mutie} I.~M.,  {del Palacio} S.,  {Beswick} R.~J.,  {Williams-Baldwin} D.,
  {Gallimore} J.~F.,  {Gallagher} J.~S.,  {Aalto} S.~E.,   {Baki} P.~O.,  2025,
  \mn@doi [\mnras] {10.1093/mnras/staf524}, \href
  {https://ui.adsabs.harvard.edu/abs/2025MNRAS.539..808M} {539, 808}

\bibitem[\protect\citeauthoryear{{Nandra} \& {Pounds}}{{Nandra} \&
  {Pounds}}{1994}]{1994MNRAS.268..405N}
{Nandra} K.,  {Pounds} K.~A.,  1994, \mn@doi [\mnras]
  {10.1093/mnras/268.2.405}, \href
  {https://ui.adsabs.harvard.edu/abs/1994MNRAS.268..405N} {268, 405}

\bibitem[\protect\citeauthoryear{{Narayan} \& {Yi}}{{Narayan} \&
  {Yi}}{1994}]{1994ApJ...428L..13N}
{Narayan} R.,  {Yi} I.,  1994, \mn@doi [\apjl] {10.1086/187381}, \href
  {https://ui.adsabs.harvard.edu/abs/1994ApJ...428L..13N} {428, L13}

\bibitem[\protect\citeauthoryear{{Narayan} \& {Yi}}{{Narayan} \&
  {Yi}}{1995}]{1995ApJ...452..710N}
{Narayan} R.,  {Yi} I.,  1995, \mn@doi [\apj] {10.1086/176343}, \href
  {https://ui.adsabs.harvard.edu/abs/1995ApJ...452..710N} {452, 710}

\bibitem[\protect\citeauthoryear{{Neilsen} \& {Lee}}{{Neilsen} \&
  {Lee}}{2009}]{2009Natur.458..481N}
{Neilsen} J.,  {Lee} J.~C.,  2009, \mn@doi [\nat] {10.1038/nature07680}, \href
  {https://ui.adsabs.harvard.edu/abs/2009Natur.458..481N} {458, 481}

\bibitem[\protect\citeauthoryear{{Parker}}{{Parker}}{1955}]{1955ApJ...121..491P}
{Parker} E.~N.,  1955, \mn@doi [\apj] {10.1086/146010}, \href
  {https://ui.adsabs.harvard.edu/abs/1955ApJ...121..491P} {121, 491}

\bibitem[\protect\citeauthoryear{{Parker}}{{Parker}}{1966}]{1966ApJ...145..811P}
{Parker} E.~N.,  1966, \mn@doi [\apj] {10.1086/148828}, \href
  {https://ui.adsabs.harvard.edu/abs/1966ApJ...145..811P} {145, 811}

\bibitem[\protect\citeauthoryear{{Parker}}{{Parker}}{1975}]{1975ApJ...198..205P}
{Parker} E.~N.,  1975, \mn@doi [\apj] {10.1086/153593}, \href
  {https://ui.adsabs.harvard.edu/abs/1975ApJ...198..205P} {198, 205}

\bibitem[\protect\citeauthoryear{{Parker}}{{Parker}}{1979}]{1979ApJ...230..914P}
{Parker} E.~N.,  1979, \apj, \href
  {https://ui.adsabs.harvard.edu/abs/1979ApJ...230..914P} {230, 914}

\bibitem[\protect\citeauthoryear{{Pinto}, {Middleton}  \& {Fabian}}{{Pinto}
  et~al.}{2016}]{2016Natur.533...64P}
{Pinto} C.,  {Middleton} M.~J.,   {Fabian} A.~C.,  2016, \mn@doi [\nat]
  {10.1038/nature17417}, \href
  {https://ui.adsabs.harvard.edu/abs/2016Natur.533...64P} {533, 64}

\bibitem[\protect\citeauthoryear{{Proga} \& {Kallman}}{{Proga} \&
  {Kallman}}{2004}]{2004ApJ...616..688P}
{Proga} D.,  {Kallman} T.~R.,  2004, \mn@doi [\apj] {10.1086/425117}, \href
  {https://ui.adsabs.harvard.edu/abs/2004ApJ...616..688P} {616, 688}

\bibitem[\protect\citeauthoryear{{Proga}, {Stone}  \& {Kallman}}{{Proga}
  et~al.}{2000}]{2000ApJ...543..686P}
{Proga} D.,  {Stone} J.~M.,   {Kallman} T.~R.,  2000, \mn@doi [\apj]
  {10.1086/317154}, \href
  {https://ui.adsabs.harvard.edu/abs/2000ApJ...543..686P} {543, 686}

\bibitem[\protect\citeauthoryear{{Reis}, {Fabian}, {Ross}, {Miniutti}, {Miller}
   \& {Reynolds}}{{Reis} et~al.}{2008}]{2008MNRAS.387.1489R}
{Reis} R.~C.,  {Fabian} A.~C.,  {Ross} R.~R.,  {Miniutti} G.,  {Miller} J.~M.,
   {Reynolds} C.,  2008, \mn@doi [\mnras] {10.1111/j.1365-2966.2008.13358.x},
  \href {https://ui.adsabs.harvard.edu/abs/2008MNRAS.387.1489R} {387, 1489}

\bibitem[\protect\citeauthoryear{Remillard \& McClintock}{Remillard \&
  McClintock}{2006}]{remillardXRayPropertiesBlackHole2006}
Remillard R.~A.,  McClintock J.~E.,  2006, \mn@doi [\araa]
  {10.1146/annurev.astro.44.051905.092532}, 44, 49

\bibitem[\protect\citeauthoryear{{Ricci} et~al.,}{{Ricci}
  et~al.}{2018}]{2018MNRAS.480.1819R}
{Ricci} C.,  et~al., 2018, \mn@doi [\mnras] {10.1093/mnras/sty1879}, \href
  {https://ui.adsabs.harvard.edu/abs/2018MNRAS.480.1819R} {480, 1819}

\bibitem[\protect\citeauthoryear{{Rodrigues}, {Giampaoli}, {Rodrigues},
  {Ferreira}, {Ter{\c{c}}as}  \& {Mendon{\c{c}}a}}{{Rodrigues}
  et~al.}{2022}]{2022Atoms..10...45R}
{Rodrigues} J.~D.,  {Giampaoli} R.,  {Rodrigues} J.~A.,  {Ferreira} A.~V.,
  {Ter{\c{c}}as} H.,   {Mendon{\c{c}}a} J.~T.,  2022, \mn@doi [Atoms]
  {10.3390/atoms10020045}, \href
  {https://ui.adsabs.harvard.edu/abs/2022Atoms..10...45R} {10, 45}

\bibitem[\protect\citeauthoryear{{Rybicki} \& {Lightman}}{{Rybicki} \&
  {Lightman}}{1986}]{1986rpa..book.....R}
{Rybicki} G.~B.,  {Lightman} A.~P.,  1986, {Radiative Processes in
  Astrophysics}

\bibitem[\protect\citeauthoryear{{Saade}, {Kaaret}, {Liodakis}  \&
  {Ehlert}}{{Saade} et~al.}{2024}]{2024ApJ...974..101S}
{Saade} M.~L.,  {Kaaret} P.,  {Liodakis} I.,   {Ehlert} S.~R.,  2024, \mn@doi
  [\apj] {10.3847/1538-4357/ad73a3}, \href
  {https://ui.adsabs.harvard.edu/abs/2024ApJ...974..101S} {974, 101}

\bibitem[\protect\citeauthoryear{{Shablovinskaya} et~al.,}{{Shablovinskaya}
  et~al.}{2024}]{2024A&A...690A.232S}
{Shablovinskaya} E.,  et~al., 2024, \mn@doi [\aap]
  {10.1051/0004-6361/202450133}, \href
  {https://ui.adsabs.harvard.edu/abs/2024A&A...690A.232S} {690, A232}

\bibitem[\protect\citeauthoryear{{Shakura} \& {Sunyaev}}{{Shakura} \&
  {Sunyaev}}{1973}]{ss73}
{Shakura} N.~I.,  {Sunyaev} R.~A.,  1973, \aap, \href
  {https://ui.adsabs.harvard.edu/abs/1973A&A....24..337S} {24, 337}

\bibitem[\protect\citeauthoryear{{Shakura} \& {Sunyaev}}{{Shakura} \&
  {Sunyaev}}{1976}]{1976MNRAS.175..613S}
{Shakura} N.~I.,  {Sunyaev} R.~A.,  1976, \mn@doi [\mnras]
  {10.1093/mnras/175.3.613}, \href
  {https://ui.adsabs.harvard.edu/abs/1976MNRAS.175..613S} {175, 613}

\bibitem[\protect\citeauthoryear{{Shapiro}, {Lightman}  \& {Eardley}}{{Shapiro}
  et~al.}{1976}]{1976ApJ...204..187S}
{Shapiro} S.~L.,  {Lightman} A.~P.,   {Eardley} D.~M.,  1976, \mn@doi [\apj]
  {10.1086/154162}, \href
  {https://ui.adsabs.harvard.edu/abs/1976ApJ...204..187S} {204, 187}

\bibitem[\protect\citeauthoryear{{Shemmer}, {Brandt}, {Netzer}, {Maiolino}  \&
  {Kaspi}}{{Shemmer} et~al.}{2006}]{2006ApJ...646L..29S}
{Shemmer} O.,  {Brandt} W.~N.,  {Netzer} H.,  {Maiolino} R.,   {Kaspi} S.,
  2006, \mn@doi [\apjl] {10.1086/506911}, \href
  {https://ui.adsabs.harvard.edu/abs/2006ApJ...646L..29S} {646, L29}

\bibitem[\protect\citeauthoryear{{Shemmer}, {Brandt}, {Netzer}, {Maiolino}  \&
  {Kaspi}}{{Shemmer} et~al.}{2008}]{2008ApJ...682...81S}
{Shemmer} O.,  {Brandt} W.~N.,  {Netzer} H.,  {Maiolino} R.,   {Kaspi} S.,
  2008, \mn@doi [\apj] {10.1086/588776}, \href
  {https://ui.adsabs.harvard.edu/abs/2008ApJ...682...81S} {682, 81}

\bibitem[\protect\citeauthoryear{{S{\k{a}}dowski}}{{S{\k{a}}dowski}}{2016}]{2016MNRAS.459.4397S}
{S{\k{a}}dowski} A.,  2016, \mn@doi [\mnras] {10.1093/mnras/stw913}, \href
  {https://ui.adsabs.harvard.edu/abs/2016MNRAS.459.4397S} {459, 4397}

\bibitem[\protect\citeauthoryear{{Stella} \& {Rosner}}{{Stella} \&
  {Rosner}}{1984}]{1984ApJ...277..312S}
{Stella} L.,  {Rosner} R.,  1984, \mn@doi [\apj] {10.1086/161697}, \href
  {https://ui.adsabs.harvard.edu/abs/1984ApJ...277..312S} {277, 312}

\bibitem[\protect\citeauthoryear{{Stepney} \& {Guilbert}}{{Stepney} \&
  {Guilbert}}{1983}]{1983MNRAS.204.1269S}
{Stepney} S.,  {Guilbert} P.~W.,  1983, \mn@doi [\mnras]
  {10.1093/mnras/204.4.1269}, \href
  {https://ui.adsabs.harvard.edu/abs/1983MNRAS.204.1269S} {204, 1269}

\bibitem[\protect\citeauthoryear{{Stern}, {Poutanen}, {Svensson}, {Sikora}  \&
  {Begelman}}{{Stern} et~al.}{1995}]{1995ApJ...449L..13S}
{Stern} B.~E.,  {Poutanen} J.,  {Svensson} R.,  {Sikora} M.,   {Begelman}
  M.~C.,  1995, \mn@doi [\apjl] {10.1086/309617}, \href
  {https://ui.adsabs.harvard.edu/abs/1995ApJ...449L..13S} {449, L13}

\bibitem[\protect\citeauthoryear{{Sunyaev} \& {Titarchuk}}{{Sunyaev} \&
  {Titarchuk}}{1980}]{1980A&A....86..121S}
{Sunyaev} R.~A.,  {Titarchuk} L.~G.,  1980, \aap, \href
  {https://ui.adsabs.harvard.edu/abs/1980A&A....86..121S} {86, 121}

\bibitem[\protect\citeauthoryear{{Suzuki} \& {Inutsuka}}{{Suzuki} \&
  {Inutsuka}}{2009}]{2009ApJ...691L..49S}
{Suzuki} T.~K.,  {Inutsuka} S.-i.,  2009, \mn@doi [\apjl]
  {10.1088/0004-637X/691/1/L49}, \href
  {https://ui.adsabs.harvard.edu/abs/2009ApJ...691L..49S} {691, L49}

\bibitem[\protect\citeauthoryear{{Suzuki}, {Muto}  \& {Inutsuka}}{{Suzuki}
  et~al.}{2010}]{2010ApJ...718.1289S}
{Suzuki} T.~K.,  {Muto} T.,   {Inutsuka} S.-i.,  2010, \mn@doi [\apj]
  {10.1088/0004-637X/718/2/1289}, \href
  {https://ui.adsabs.harvard.edu/abs/2010ApJ...718.1289S} {718, 1289}

\bibitem[\protect\citeauthoryear{{Svensson} \& {Zdziarski}}{{Svensson} \&
  {Zdziarski}}{1994}]{1994ApJ...436..599S}
{Svensson} R.,  {Zdziarski} A.~A.,  1994, \mn@doi [\apj] {10.1086/174934},
  \href {https://ui.adsabs.harvard.edu/abs/1994ApJ...436..599S} {436, 599}

\bibitem[\protect\citeauthoryear{{Taam} \& {Lin}}{{Taam} \&
  {Lin}}{1984}]{1984ApJ...287..761T}
{Taam} R.~E.,  {Lin} D.~N.~C.,  1984, \mn@doi [\apj] {10.1086/162734}, \href
  {https://ui.adsabs.harvard.edu/abs/1984ApJ...287..761T} {287, 761}

\bibitem[\protect\citeauthoryear{{Tabone}, {Rosotti}, {Cridland}, {Armitage}
  \& {Lodato}}{{Tabone} et~al.}{2022}]{2022MNRAS.512.2290T}
{Tabone} B.,  {Rosotti} G.~P.,  {Cridland} A.~J.,  {Armitage} P.~J.,   {Lodato}
  G.,  2022, \mn@doi [\mnras] {10.1093/mnras/stab3442}, \href
  {https://ui.adsabs.harvard.edu/abs/2022MNRAS.512.2290T} {512, 2290}

\bibitem[\protect\citeauthoryear{{Tamilan}, {Hayasaki}  \& {Suzuki}}{{Tamilan}
  et~al.}{2025}]{2025PTEP.2025b3E02T}
{Tamilan} M.,  {Hayasaki} K.,   {Suzuki} T.~K.,  2025, \mn@doi [Progress of
  Theoretical and Experimental Physics] {10.1093/ptep/ptaf014}, \href
  {https://ui.adsabs.harvard.edu/abs/2025PTEP.2025b3E02T} {2025, 023E02}

\bibitem[\protect\citeauthoryear{{Titarchuk}}{{Titarchuk}}{1994}]{1994ApJ...434..570T}
{Titarchuk} L.,  1994, \mn@doi [\apj] {10.1086/174760}, \href
  {https://ui.adsabs.harvard.edu/abs/1994ApJ...434..570T} {434, 570}

\bibitem[\protect\citeauthoryear{{Tombesi}, {Cappi}, {Reeves}, {Palumbo},
  {Yaqoob}, {Braito}  \& {Dadina}}{{Tombesi}
  et~al.}{2010}]{2010A&A...521A..57T}
{Tombesi} F.,  {Cappi} M.,  {Reeves} J.~N.,  {Palumbo} G.~G.~C.,  {Yaqoob} T.,
  {Braito} V.,   {Dadina} M.,  2010, \mn@doi [\aap]
  {10.1051/0004-6361/200913440}, \href
  {https://ui.adsabs.harvard.edu/abs/2010A&A...521A..57T} {521, A57}

\bibitem[\protect\citeauthoryear{{Tombesi}, {Cappi}, {Reeves}  \&
  {Braito}}{{Tombesi} et~al.}{2012}]{2012MNRAS.422L...1T}
{Tombesi} F.,  {Cappi} M.,  {Reeves} J.~N.,   {Braito} V.,  2012, \mn@doi
  [\mnras] {10.1111/j.1745-3933.2012.01221.x}, \href
  {https://ui.adsabs.harvard.edu/abs/2012MNRAS.422L...1T} {422, L1}

\bibitem[\protect\citeauthoryear{{Tombesi}, {Cappi}, {Reeves}, {Nemmen},
  {Braito}, {Gaspari}  \& {Reynolds}}{{Tombesi}
  et~al.}{2013}]{2013MNRAS.430.1102T}
{Tombesi} F.,  {Cappi} M.,  {Reeves} J.~N.,  {Nemmen} R.~S.,  {Braito} V.,
  {Gaspari} M.,   {Reynolds} C.~S.,  2013, \mn@doi [\mnras]
  {10.1093/mnras/sts692}, \href
  {https://ui.adsabs.harvard.edu/abs/2013MNRAS.430.1102T} {430, 1102}

\bibitem[\protect\citeauthoryear{{Turner}, {Blaes}, {Socrates}, {Begelman}  \&
  {Davis}}{{Turner} et~al.}{2005}]{2005ApJ...624..267T}
{Turner} N.~J.,  {Blaes} O.~M.,  {Socrates} A.,  {Begelman} M.~C.,   {Davis}
  S.~W.,  2005, \mn@doi [\apj] {10.1086/428723}, \href
  {https://ui.adsabs.harvard.edu/abs/2005ApJ...624..267T} {624, 267}

\bibitem[\protect\citeauthoryear{{Uttley}, {Cackett}, {Fabian}, {Kara}  \&
  {Wilkins}}{{Uttley} et~al.}{2014}]{2014A&ARv..22...72U}
{Uttley} P.,  {Cackett} E.~M.,  {Fabian} A.~C.,  {Kara} E.,   {Wilkins} D.~R.,
  2014, \mn@doi [\aapr] {10.1007/s00159-014-0072-0}, \href
  {https://ui.adsabs.harvard.edu/abs/2014A&ARv..22...72U} {22, 72}

\bibitem[\protect\citeauthoryear{{Vasudevan} \& {Fabian}}{{Vasudevan} \&
  {Fabian}}{2007}]{2007MNRAS.381.1235V}
{Vasudevan} R.~V.,  {Fabian} A.~C.,  2007, \mn@doi [\mnras]
  {10.1111/j.1365-2966.2007.12328.x}, \href
  {https://ui.adsabs.harvard.edu/abs/2007MNRAS.381.1235V} {381, 1235}

\bibitem[\protect\citeauthoryear{{Weisskopf} et~al.,}{{Weisskopf}
  et~al.}{2022}]{2022JATIS...8b6002W}
{Weisskopf} M.~C.,  et~al., 2022, \mn@doi [Journal of Astronomical Telescopes,
  Instruments, and Systems] {10.1117/1.JATIS.8.2.026002}, \href
  {https://ui.adsabs.harvard.edu/abs/2022JATIS...8b6002W} {8, 026002}

\bibitem[\protect\citeauthoryear{Wieselsberger}{Wieselsberger}{1922}]{wieselsberger1922new}
Wieselsberger C.,  1922, Technical report, New data on the laws of fluid
  resistance

\bibitem[\protect\citeauthoryear{{Xu}, {Cao}, {Wang}  \& {Zdziarski}}{{Xu}
  et~al.}{2025}]{2025MNRAS.544.1748X}
{Xu} H.,  {Cao} X.,  {Wang} Y.,   {Zdziarski} A.~A.,  2025, \mn@doi [\mnras]
  {10.1093/mnras/staf1877}, \href
  {https://ui.adsabs.harvard.edu/abs/2025MNRAS.544.1748X} {544, 1748}

\bibitem[\protect\citeauthoryear{{Yamada}, {Yoo}, {Jara-Almonte}, {Ji},
  {Kulsrud}  \& {Myers}}{{Yamada} et~al.}{2014}]{2014NatCo...5.4774Y}
{Yamada} M.,  {Yoo} J.,  {Jara-Almonte} J.,  {Ji} H.,  {Kulsrud} R.~M.,
  {Myers} C.~E.,  2014, \mn@doi [Nature Communications] {10.1038/ncomms5774},
  \href {https://ui.adsabs.harvard.edu/abs/2014NatCo...5.4774Y} {5, 4774}

\bibitem[\protect\citeauthoryear{{You}, {Cao}  \& {Yuan}}{{You}
  et~al.}{2012}]{2012ApJ...761..109Y}
{You} B.,  {Cao} X.,   {Yuan} Y.-F.,  2012, \mn@doi [\apj]
  {10.1088/0004-637X/761/2/109}, \href
  {https://ui.adsabs.harvard.edu/abs/2012ApJ...761..109Y} {761, 109}

\bibitem[\protect\citeauthoryear{{Zdziarski}}{{Zdziarski}}{1998}]{1998MNRAS.296L..51Z}
{Zdziarski} A.~A.,  1998, \mn@doi [\mnras] {10.1046/j.1365-8711.1998.01682.x},
  \href {https://ui.adsabs.harvard.edu/abs/1998MNRAS.296L..51Z} {296, L51}

\bibitem[\protect\citeauthoryear{{Zdziarski}, {Johnson}  \&
  {Magdziarz}}{{Zdziarski} et~al.}{1996}]{1996MNRAS.283..193Z}
{Zdziarski} A.~A.,  {Johnson} W.~N.,   {Magdziarz} P.,  1996, \mn@doi [\mnras]
  {10.1093/mnras/283.1.193}, \href
  {https://ui.adsabs.harvard.edu/abs/1996MNRAS.283..193Z} {283, 193}

\bibitem[\protect\citeauthoryear{{Zhang}, {Blaes}  \& {Jiang}}{{Zhang}
  et~al.}{2021}]{2021MNRAS.508..617Z}
{Zhang} L.,  {Blaes} O.,   {Jiang} Y.-F.,  2021, \mn@doi [\mnras]
  {10.1093/mnras/stab2510}, \href
  {https://ui.adsabs.harvard.edu/abs/2021MNRAS.508..617Z} {508, 617}

\bibitem[\protect\citeauthoryear{{Zhang} et~al.,}{{Zhang}
  et~al.}{2022a}]{2022MNRAS.512.2686Z}
{Zhang} Y.,  et~al., 2022a, \mn@doi [\mnras] {10.1093/mnras/stac690}, \href
  {https://ui.adsabs.harvard.edu/abs/2022MNRAS.512.2686Z} {512, 2686}

\bibitem[\protect\citeauthoryear{{Zhang}, {Blaes}  \& {Jiang}}{{Zhang}
  et~al.}{2022b}]{2022MNRAS.515.4371Z}
{Zhang} L.,  {Blaes} O.,   {Jiang} Y.-F.,  2022b, \mn@doi [\mnras]
  {10.1093/mnras/stac1815}, \href
  {https://ui.adsabs.harvard.edu/abs/2022MNRAS.515.4371Z} {515, 4371}

\bibitem[\protect\citeauthoryear{{Zhang} et~al.,}{{Zhang}
  et~al.}{2025}]{2025PASJ..tmp..125Z}
{Zhang} S.,  et~al., 2025, \mn@doi [\pasj] {10.1093/pasj/psaf127}, \href
  {https://ui.adsabs.harvard.edu/abs/2025PASJ..tmp..125Z} {}

\bibitem[\protect\citeauthoryear{{Zhang} et~al.,}{{Zhang}
  et~al.}{2026}]{2026MNRAS.546ag261Z}
{Zhang} Y.,  et~al., 2026, \mn@doi [\mnras] {10.1093/mnras/stag261}, \href
  {https://ui.adsabs.harvard.edu/abs/2026MNRAS.546ag261Z} {546, stag261}

\bibitem[\protect\citeauthoryear{{Zhu} \& {Stone}}{{Zhu} \&
  {Stone}}{2018}]{2018ApJ...857...34Z}
{Zhu} Z.,  {Stone} J.~M.,  2018, \mn@doi [\apj] {10.3847/1538-4357/aaafc9},
  \href {https://ui.adsabs.harvard.edu/abs/2018ApJ...857...34Z} {857, 34}

\makeatother
\end{thebibliography}







	\bsp 
	\label{lastpage}
\end{document}